\documentclass[aps,pre,showpacs,amsmath,amssymb,amsfonts,superscriptaddress,onecolumn,longbibliography]{revtex4-1}

\usepackage{graphicx}
\usepackage{subfigure}
\usepackage{verbatim}
\usepackage{dcolumn}
\usepackage{bm}
\usepackage{epsf}
\usepackage{color}
\usepackage[colorlinks=true,citecolor=blue,linkcolor=blue]{hyperref}

\newcommand{\bla}{bla\\bla\\bla\\bla\\bla}

\newcommand{\PRL}{Phys. Rev. Lett. }

\definecolor{amethyst}{rgb}{0.6, 0.4, 0.8}
\definecolor{asparagus}{rgb}{0.53, 0.66, 0.42}
\definecolor{fuzzywuzzy}{rgb}{0.8, 0.4, 0.4}

\begin{document}

\title{The quantum-classical correspondence principle for work distributions}

\author{Christopher Jarzynski}
\email{cjarzyns@umd.edu}
\affiliation{Department of Chemistry and Biochemistry and Institute of Physical Science and Technology, University of Maryland, College Park, MD 20742, U.S.A.}

\author{H. T. Quan}
\email{htquan@pku.edu.cn}
\affiliation{School of Physics, Peking University, Beijing 100871, China}
\affiliation{Collaborative Innovation Center of Quantum Matter, Beijing 100871, China}

\author{Saar Rahav}
\email{rahavs@tx.technion.ac.il}
\affiliation{Schulich Faculty of Chemistry, Technion-Israel Institute of Technology, Haifa 32000, Israel}

\date{\today}

\begin{abstract}
For closed quantum systems driven away from equilibrium, {\it work} is often defined in terms of projective measurements of initial and final energies.
This definition leads to statistical distributions of work that satisfy nonequilibrium work and fluctuation relations.
While this two-point measurement definition of quantum work can be justified heuristically by appeal to the first law of thermodynamics, its relationship to the classical definition of work has not been carefully examined.
In this paper we employ semiclassical methods, combined with numerical simulations of a driven quartic oscillator, to study the correspondence between classical and quantal definitions of work in systems with one degree of freedom.
We find that a semiclassical work distribution, built from classical trajectories that connect the initial and final energies, provides an excellent approximation to the quantum work distribution when the trajectories are assigned suitable phases and are allowed to interfere. Neglecting the interferences between trajectories reduces the distribution to that of the corresponding classical process.
Hence, in the semiclassical limit, the quantum work distribution converges to the classical distribution, decorated by a quantum interference pattern.
We also derive the form of the quantum work distribution at the boundary between classically allowed and forbidden regions, where this distribution tunnels into the forbidden region.
Our results clarify how the correspondence principle applies in the context of quantum and classical work distributions, and contribute to the understanding of work and nonequilibrium work relations in the quantum regime.
\end{abstract}

 \pacs{05.70.Ln, 
 05.30.-d, 
    05.90.+m. 
}

\maketitle


\section{Introduction}
\label{sec:intro}

{\it Work} is a familiar concept in elementary mechanics and a central one in thermodynamics.
In recent years, interest in the nonequilibrium thermodynamics of small systems~\cite{Bustamante2005,Jarzynski2011,Seifert2012,Klagesbook} has motivated careful examinations of how to define {\it quantum work}~\cite{Piechocinska2000,Kurchan2000,Tasaki2000,Yukawa2000,Mukamel2003,Monnai2003,Chernyak2004,DeRoeck2004,Allahverdyan2005,Engel2007,Talkner2007,Teifel2007,Crooks2008,Campisi2009,Esposito2009,Campisi2011,Campisi2011pre,Horowitz2012,Liu2012,Subasi2012,Albash2013,Talkner2013,Deffner2013_epl,Hekking2013,Solinas2013,Liu2014a,Liu2014b,Silaev2014,Salmilehto2014,Rastegin2014,Watanabe2014,Roncaglia2014,Allahverdyan2014,Gong2014,Halpern2014,Suomela2014,Viisanen2014,Jennings2015,Hanggi2015,Talkner2015,Suomela2015,Solinas2015,Goold2015}.
In this context one often considers a process in which a quantum system evolves under the Schr\" odinger equation as its Hamiltonian is varied in time -- for instance a quantum particle in a piston undergoing compression or expansion~\cite{Quan2012}.
It is typically assumed that the system is initialized in thermal equilibrium, and the question becomes: how do we appropriately define the work performed on the system during a single realization of this process?

One answer involves two projective measurements of the system's energy, at the start of the process ($t=0$) and at the end ($t=\tau$).
If the system Hamiltonian is varied from $\hat H(0) = \hat H_A$ to $\hat H(\tau) = \hat H_B$, then the measurement outcomes will be eigenvalues of these operators.
The work performed during the process is then defined to be the difference between these two values, e.g.
\begin{equation}
\label{eq:qwork}
W = E_n^B - E_m^A
\end{equation}
if the measurements produce the $m$'th and $n$'th eigenvalues of the initial and final Hamiltonians.
In this definition, work is inherently stochastic, with
two sources of randomness: the statistical randomness associated with sampling an initial energy $E_m^A$ from the canonical equilibrium distribution (Eq.~\ref{eq:pqm}), and the purely quantal randomness associated with the ``collapse'' of the final wavefunction $\vert\psi_\tau\rangle$ upon making a projective measurement of the final energy (Eq.~\ref{eq:Pnm})~\cite{Neumann1955}.

In an analogous classical process, a system is prepared in thermal equilibrium, then it evolves under Hamilton's equations as the Hamiltonian function is varied from $H_A(z)$ to $H_B(z)$, where $z$ denotes a point in the system's phase space.
Since the system is thermally isolated, it is natural to define the work as the change in its internal energy:
\begin{equation}
\label{eq:cwork}
W = H_B(z_\tau) - H_A(z_0)
\end{equation}
for a realization during which the system evolves from $z_0$ to $z_\tau$.
As in the quantum case $W$ is a stochastic variable, but here the randomness arises solely from the sampling of the initial microstate $z_0$ from an equilibrium distribution.

Much of the recent interest in quantum work has been stimulated by the discovery and experimental verification of classical {\it nonequilibrium work relations}~\cite{Bochkov1977,Jarzynski1997a,Crooks1999,Hummer2001,Liphardt2002,Collin2005}, which have provided insights into the second law of thermodynamics, particularly as it applies to small systems where fluctuations are important~\cite{Jarzynski2011}.
For thermally isolated systems, the quantal counterparts of these relations follow directly from the definition of work given by Eq.~\ref{eq:qwork}~\cite{Kurchan2000,Tasaki2000,Mukamel2003,Talkner2007}.
Moreover, experimental tests of quantum nonequilibrium work relations have recently been proposed~\cite{Huber2008,Dorner2013,Mazzola2013,Campisi2013} and implemented~\cite{Batalhao2014,An2015}, providing direct verification of the validity of these results.

Despite the evident similarity between Eqs.~\ref{eq:qwork} and \ref{eq:cwork}, and despite the relevance of Eq.~\ref{eq:qwork} to nonequilibrium work relations, the definition of quantum work given by Eq.~\ref{eq:qwork} might seem {\it ad hoc}.
Indeed, given the absence of a broadly agreed-upon ``textbook'' definition of quantum work, one might reasonably suspect that Eq.~\ref{eq:qwork} has been introduced precisely because it leads to quantum nonequilibrium work relations.
Furthermore, the wavefunction collapse that occurs upon measuring the final energy has no classical counterpart.
Due to this collapse, the quantum work has no independent physical reality until the measurement is performed at $t=\tau$.
By contrast the classical work can be viewed as a well-defined function of time, namely the net change in the value of the time-dependent Hamiltonian along the trajectory $z_t$.
Our aim in this paper is to use the tools of semiclassical mechanics, together with numerical simulations, to investigate the relationship between Eqs.~\ref{eq:qwork} and \ref{eq:cwork}, and specifically to clarify how the correspondence principle applies in this context.

We restrict ourselves to systems with one degree of freedom, for which there exist explicit semiclassical approximations of energy eigenstates.
We focus on the transition probability $P^Q(n \vert m)$ from the $m$'th eigenstate of $\hat H_A$ to the $n$'th eigenstate of $\hat H_B$, and on its classical analogue, $P^C(n \vert m)$, defined in Sec.~\ref{sec:setup}.
In Sec.~\ref{sec:numerical} we investigate both quantities numerically for the example of a forced quartic oscillator.
For high-lying initial energies $P^Q$ oscillates rapidly with $n$, whereas $P^C$ is a smooth function over a finite range $n_{\rm min} < n < n_{\rm max}$ (Fig.~\ref{comparison}).
After integrating over the oscillations the two quantities are nearly identical (Fig.~\ref{transitionprobabilities}), which provides some justification for viewing Eq.~\ref{eq:qwork} as the quantal counterpart of Eq.~\ref{eq:cwork}.
In Sec.~\ref{sec:semiclassical} we derive $P^{SC}(n \vert m)$, a semiclassical approximation for $P^Q$ expressed as a sum over classical trajectories.
Each trajectory in this sum carries a quantal phase, giving rise to coherent interferences between trajectories.
When these interferences are neglected, $P^{SC}$ agrees with $P^C$; when they are included, $P^{SC}$ accurately captures the oscillations in $P^Q$ (Fig.~\ref{sctransitionprobabilities}).
Thus the oscillations in $P^Q$ can be understood as a quantum interference pattern superimposed on a classical background.
This picture breaks down around $n_{\rm min}$ and $n_{\rm max}$, where $P^Q$ exhibits tails that tunnel into classically forbidden regions.
In Sec.~\ref{sec:airy} we derive a semiclassical approximation, expressed in terms of the Airy function, that accurately describes these tails (Fig.~\ref{perfect}).

The analysis in Secs.~\ref{sec:semiclassical} and \ref{sec:airy} relies on theoretical tools that are familiar in the field of semiclassical mechanics.
To the best of our knowledge, these tools have not previously been applied to study the relationship between classical and quantum work distributions, although similar analyses have been performed in the context of molecular scattering theory~\cite{Marcus1970,Marcus1971,Miller1974} and laser-pulsed atoms~\cite{Schwieters1995a,Schwieters1995b}, among other examples.
In particular, our calculations in Sec.~\ref{subsec:sctp} closely parallel those of Schwieters and Delos~\cite{Schwieters1995b}.

\section{Classical and quantum transition probabilities}
\label{sec:setup}

Here we introduce notation and specify the problem we plan to study.
In Sec. \ref{subsec:csetup} we define a discretized classical work distribution, Eq.~\ref{eq:cworkDist_sum}, that can be compared directly to the quantum work distribution, Eq.~\ref{eq:qworkDist}.
These distributions involve classical and quantum transition probabilities, $P^C(n \vert m)$ and $P^Q(n \vert m)$, which will be the central objects of study throughout the rest of the paper.

\subsection{Classical setup}
\label{subsec:csetup}

Consider a system with one degree of freedom, described by a Hamiltonian
\begin{equation}
\label{eq:hamiltonian}
H(z;\lambda) = \frac{p^2}{2M} + V(q;\lambda) \quad ,
\end{equation}
where $z=(q,p)$ denotes a point in phase space, and $\lambda$ is an externally controlled parameter.
We assume that the energy shells (level surfaces) of the Hamiltonian form simple, closed curves in phase space.
We will consider the evolution of this system under the time-dependent Hamiltonian $H(z;\lambda_t)$, where $\lambda$ is varied from $\lambda_0=A$ to $\lambda_\tau=B$.
For compact notation, we define $H_A(z) \equiv H(z;\lambda_0)$ and $H_B(z) \equiv H(z;\lambda_\tau)$.

Suppose that prior to the start of the process, the system has come to equilibrium with a thermal reservoir at temperature $T$ and the reservoir has been removed.
Therefore at $t=0$ the microstate of the system will be treated as a random sample from a canonical distribution corresponding to the Hamiltonian $H_A$.
From $t=0$ to $t=\tau$, the system is described by a trajectory $z_t$ evolving under Hamilton's equations, and the work performed on the system is given by the difference between its initial and final energies, as per Eq.~\ref{eq:cwork}:
\begin{equation}
W = E_\tau - E_0 \equiv  H_B(z_\tau) - H_A(z_0) \quad .
\end{equation}
For an ensemble of realizations of this process, the distribution of values of work performed on the system is
\begin{equation}
\label{eq:cworkDist}
P^C(W) = \int {\rm d}E_\tau \int {\rm d}E_0 \, \bar P^C(E_\tau \vert E_0) \, \bar P_A^C(E_0) \, \delta(W - E_\tau + E_0)  \quad ,
\end{equation}
where $\bar P_A^C(E_0)$ is the probability distribution of initial energies, sampled from equilibrium, and $\bar P^C(E_\tau \vert E_0)$ is the conditional probability distribution to end with a final energy $E_\tau$, given an initial energy $E_0$.
Let us consider these factors separately.

$\bar P_A^C$ is simply the classical equilibrium energy distribution:
\begin{equation}
\label{eq:classicalEDist}
\bar P_A^C(E_0) = \frac{1}{Z_A^C} e^{-\beta E_0} g_A(E_0) \quad,
\end{equation}
where
\begin{equation}
Z_\lambda^C = \frac{1}{h} \int {\rm d}z \, \exp\left[ -\beta H(z;\lambda) \right]
\qquad \textrm{and} \qquad
g_\lambda(E) = \frac{1}{h} \int {\rm d}z \, \delta\left[ E - H(z;\lambda) \right]
\end{equation}
are the classical partition function and density of states, $\beta = (k_BT)^{-1}$, and $h$ is Planck's constant.
We have included the factors $h^{-1}$ (which cancel in Eq.~\ref{eq:classicalEDist}) for later convenience.
Let us also define
\begin{equation}
\Omega(E,\lambda) = \int {\rm d}z \, \theta\left[ E - H(z;\lambda) \right] = \oint_E p \, {\rm d}q \quad ,
\end{equation}
which is the phase space volume enclosed by the energy shell $E$ of the Hamiltonian $H$, equivalently the integral of $p \, {\rm d}q$ around this energy shell.
We then have
\begin{equation}
\label{eq:cdos}
g_\lambda(E) = \frac{1}{h} \frac{\partial\Omega}{\partial E}
\end{equation}

\begin{figure}[tbp]
	\subfigure[\ $t=0$]{\label{fig:A0}
	\includegraphics[trim = 1in 0in 0in 0in , scale=0.3,angle=0]{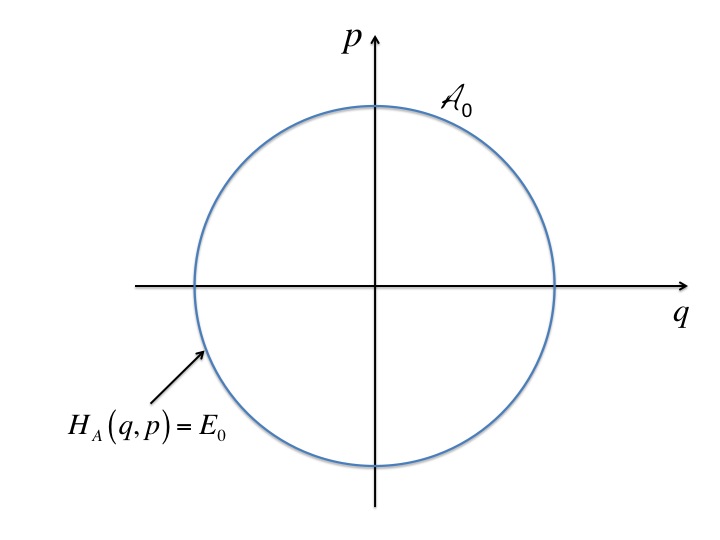}
	}
	\subfigure[\ $t=\tau$]{\label{Atau_B}
	\includegraphics[trim = 0in 0in 0in 0in , scale=0.3,angle=0]{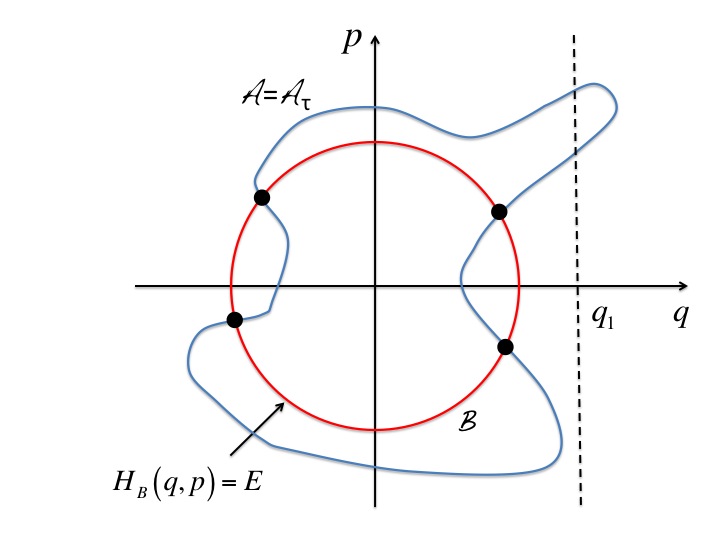}
	}
\caption{An energy shell ${\cal A}_0$ of the initial Hamiltonian evolves under $H(t)$ to the curve ${\cal A}\equiv{\cal A}_\tau$.
The curve ${\cal B}$ is an energy shell of the final Hamiltonian.}
\label{fig:LagrangianManifolds}
\end{figure}

The conditional probability distribution $\bar P^C(E_\tau \vert E_0)$ is given by the expression
\begin{equation}
\label{eq:PEE0}
\bar P^C(E_\tau \vert E_0) = \frac{\int {\rm d}z_{0} \ \delta\left[ E_0 - H_A(z_{0}) \right] \, \delta\left[ E_\tau - H_B(z_\tau(z_{0})) \right]}{\int {\rm d}z_{0} \ \delta\left[ E_0 - H_A(z_{0}) \right]}
\quad,
\end{equation}
where $z_\tau(z_0)$ denotes the final conditions of a trajectory that evolves from initial conditions $z_0$.
It is useful to imagine an ensemble of initial conditions sampled microcanonically from the energy shell $E_0$ of the Hamiltonian $H_A$ (see Fig.~\ref{fig:A0}).
The swarm of trajectories that evolves from these initial conditions defines a time-dependent, closed curve ${\cal A}_t$ in phase space.
At $t=\tau$, the points of intersection between ${\cal A}_\tau$ and the energy shell $E_\tau$ of the Hamiltonian $H_B$ (black dots in Fig.~\ref{Atau_B}) represent the final conditions of those trajectories that end with energy $E_\tau$, having begun with energy $E_0$.
If we consider two nearby energy shells $E_\tau$ and $E_\tau + {\rm d} E$, then $\bar P^C(E_\tau \vert E_0) \, {\rm d} E_\tau$ is the fraction of trajectories whose final energies fall within this energy interval.

Let us define the {\it m'th energy interval} of the Hamiltonian $H_A$ to be the range of energy values $E$ satisfying
\begin{equation}
\label{eq:minterval}
mh \le \Omega(E,A) < (m+1)h
\end{equation}
where $m=0,1,2 \cdots$.
Furthermore, let us use a mid-point rule to assign a particular energy value $E_m^A$ to each interval:
\begin{equation}
\label{eq:midpoint}
\Omega(E_m^A,A) = \oint_{E_m^A} p \, {\rm d}q = \left( m + \frac{1}{2} \right) h \quad .
\end{equation}
Analogous definitions apply to the $n$'th energy interval of the Hamiltonian $H_B$, and the corresponding energy $E_n^B$.

From Eqs.~\ref{eq:cdos} and \ref{eq:minterval} the width of the $m$'th energy interval of $H_A$ is
\begin{equation}
\label{eq:mwidth}
\delta E_m^A \approx \frac{1}{g_A(E_m^A)} \quad.
\end{equation}
For a smooth function of energy $f(E)$, Eq.~\ref{eq:mwidth} leads to
\begin{equation}
\label{eq:intApprox}
\int_m {\rm d}E_0 f(E_0) \approx \frac{f(E_m^A)}{g_A(E_m^A)}  \quad,
\end{equation}
where the integral is taken over the $m$'th interval.

Readers will recognize Eq.~\ref{eq:midpoint} as the semiclassical quantization condition $\oint p\,{\rm d}q = [m+(1/2)] h$, which provides an excellent approximation for the $m$'th eigenvalue of the quantum Hamiltonian $\hat H_A$ (Eq.~\ref{eq:hatH} below) when $m\gg 1$.
For convenience, we will later use the notation $E_m^A$ to denote this eigenvalue, without making a distinction between the exact eigenvalue and its semiclassical approximation.

The probability to obtain initial conditions $z_0$ within the $m$'th interval of $H_A$, when sampling from equilibrium, is
\begin{equation}
\label{eq:pcm}
P_A^C(m) = \int_m {\rm d}E_0 \, \bar P_A^C(E_0) \approx \frac{1}{Z_A^C} e^{-\beta E_m^A} \quad,
\end{equation}
using Eq.~\ref{eq:intApprox}.
Similarly, the conditional probability to end in the $n$'th energy interval, given a representative initial energy in the $m$'th interval, is
\begin{equation}
\label{eq:pcnm}
P^C(n \vert m) = \int_n {\rm d}E_\tau  \, \bar P^C(E_\tau \vert E_0 = E_m^A) \approx \frac{\bar P^C(E_n^B \vert E_m^A)}{g_B(E_n^B)}   \quad .
\end{equation}
We will refer to this as the {\it classical transition probability}.
The interpretation of $P^C(n \vert m)$ is straightforward. Imagine a swarm of trajectories evolving from initial conditions sampled from a microcanonical ensemble with energy $E_{m}^{A}$ (see Fig.~\ref{fig:A0}).
At the end of the process, $t=\tau$,
the fraction of these trajectories that fall into the energy window $[E_{n}^{B}, E_{n+1}^{B} ]$ is equal to $P^C(n \vert m)$ \cite{Schwieters1995b}.
The classical work distribution can now be rewritten as a sum over initial and final energy intervals:
\begin{equation}
\label{eq:cworkDist_sum}
\begin{split}
P^C(W) &= \sum_{n,m}  \int_n {\rm d}E_\tau \int_m {\rm d}E_0 \, \bar P^C(E_\tau \vert E_0) \, \bar P_A^C(E_0) \, \delta(W - E_\tau + E_0) \\
&\approx \sum_{n,m}  \int_n {\rm d}E_\tau \, \int_m {\rm d}E_0 \, \bar P^C(E_\tau \vert E_m^A)  \, \bar P_A^C(E_0) \, \delta(W - E_n^B + E_m^A) \\
&= \sum_{m,n} P^C(n \vert m) \, P_A^C(m) \, \delta(W - E_n^B + E_m^A) \quad .
\end{split}
\end{equation}

Finally, we point out that the approximations appearing in Eqs.~\ref{eq:mwidth} - \ref{eq:cworkDist_sum} assume that the width of the $m$'th energy interval, $\delta E_m$, is very small on a classical scale.
Formally, this assumption can be justified by taking the semiclassical limit, $\hbar \rightarrow 0$, with all classical quantities held fixed.
In this limit the density of states increases, as does the quantum number at a given energy.
In practice the semiclassical limit is often simply identified with the condition $m\gg 1$.

\subsection{Quantum setup}

In the quantum version of this problem, $H(z;\lambda)$ is replaced by the Hermitian operator
\begin{equation}
\label{eq:hatH}
\hat H(\lambda) = -\frac{\hbar^2}{2M} \frac{\partial^2}{\partial q^2} + V(q;\lambda) \quad .
\end{equation}
A wavefunction $\psi(q,t) = \langle q \vert \psi_t \rangle$ evolves under the Schr\" odinger equation 
as $\hat H(\lambda_t)$ is varied from $\hat H(\lambda_0)=\hat H_A$ to $\hat H(\lambda_\tau)=\hat H_B$.
The eigenstates and eigenvalues of $\hat H_A$ are specified as follows:
\begin{equation}
\phi_m^A(q) = \langle q \vert \phi_m^A\rangle \quad , \quad \hat H_A \vert\phi_m^A\rangle = E_m^A  \vert\phi_m^A\rangle
\label{eigenstate}
\end{equation}
with similar notation for $\hat H_B$.
Evolution in time is represented by the unitary operator $\hat U_t$ satisfying
$\hat H(\lambda_t) \hat U_t = i\hbar\, \partial \hat U_t / \partial t$ and $\hat U_0 = \hat I$, where $\hat I$ is the identity operator.


Using Eq.~\ref{eq:qwork} and assuming thermal equilibration at $t=0$, the work distribution is~\cite{Talkner2007}
\begin{equation}
\label{eq:qworkDist}
P^Q(W) = \sum_{n,m} P^Q(n \vert m) \, P_A^Q(m) \, \delta(W - E_n^B + E_m^A) \quad .
\end{equation}
$P_A^Q(m)$ is the probability of obtaining the $m$'th eigenstate of $\hat H_A$ when making the initial energy measurement:
\begin{equation}
\label{eq:pqm}
P_A^Q(m) = \frac{e^{-\beta E_m^A}}{Z_A^Q}
\quad,
\end{equation}
with $Z_A^Q = \sum_m \exp(-\beta E_m^A)$.
The {\it quantum transition probability} $P^Q(n \vert m)$ is the conditional probability to obtain the $n$'th eigenstate of $\hat H_B$ upon making the final measurement, given the $m$'th eigenstate of $\hat H_A$ at the initial measurement:
\begin{equation}
\label{eq:Pnm}
P^Q(n \vert m) = \left\vert \left\langle \phi_n^B \vert \hat U_\tau \vert \phi_m^A \right\rangle \right\vert^2
= \left\vert \left\langle \phi_n^B \vert  \psi_\tau \right\rangle \right\vert^2
\end{equation}

The classical and quantum work distributions given by Eqs.~\ref{eq:cworkDist_sum} and \ref{eq:qworkDist} can now be compared directly.
We note that
\begin{equation}
Z_A^C
= \int {\rm d}E \, e^{-\beta E} g_A(E) \approx \sum_m e^{-\beta E_m^A} = Z_A^Q \quad,
\end{equation}
hence $P_A^C(m) \approx P_A^Q(m)$ (see Eqs.~\ref{eq:pcm} and \ref{eq:pqm}).
Thus what remains is to clarify the relationship between the classical and quantum transition probabilities, $P^C(n \vert m)$ and $P^Q(n \vert m)$.
In the following section, we compare these transition probabilities in a model system for which both the classical and quantum dynamics are simulated numerically.
We will find that $P^C$ and $P^Q$ are manifestly different (Fig.~\ref{comparison}), but these differences mostly vanish after appropriate smoothing (Fig.~\ref{transitionprobabilities}).
In Sections~\ref{sec:semiclassical} and \ref{sec:airy} we develop a semiclassical theory to explain these features.

\section{Numerical case study: the forced quartic oscillator}
\label{sec:numerical}

We begin with a model system, the forced quartic oscillator:
\begin{equation}
H(z;\lambda) = \frac{p^2}{2M} + \lambda q^4 \quad .
\label{Hamiltonian}
\end{equation}
We set $M=1/2$ and $\hbar=h/2\pi = 1$, and we vary the work parameter at a constant rate, $\lambda_t = \lambda_0 + vt$, from $\lambda_0=1=A$ to $\lambda_\tau=5=B$, taking $v=50$ and therefore $\tau=0.08$.

A number of groups have previously compared quantum and classical work distributions for a driven {\it harmonic} oscillator, for which analytical solutions are available.
Specifically,
Deffner {\it et al}~\cite{Deffner2008,Deffner2010} have studied an oscillator with a time-varying stiffness; Talkner {\it et al}~\cite{Talkner2008} have obtained the work distribution for an oscillator driven by time-dependent perturbations that are linear in $q$ and $p$;
Campisi~\cite{Campisi2008b} has studied changes in the Boltzmann entropy for forced quantum and classical oscillators;
Ford {\it et al}~\cite{Ford2012} have investigated a harmonic oscillator with a cyclically driven stiffness; and Talkner {\it et al} have considered the sudden quench of a two-dimensional  oscillator~\cite{Talkner2013}.
While the harmonic oscillator has the advantage of being analytically tractable, it is somewhat special in that its quantum dynamics can be reduced to its classical dynamics \cite{Husimi1953}.
Here we have chosen the more ``generic'' quartic oscillator, which requires numerical simulations.

\begin{figure}[tbh]
\includegraphics[trim = 0in 0in 0in 0in , scale=1.0,angle=0]{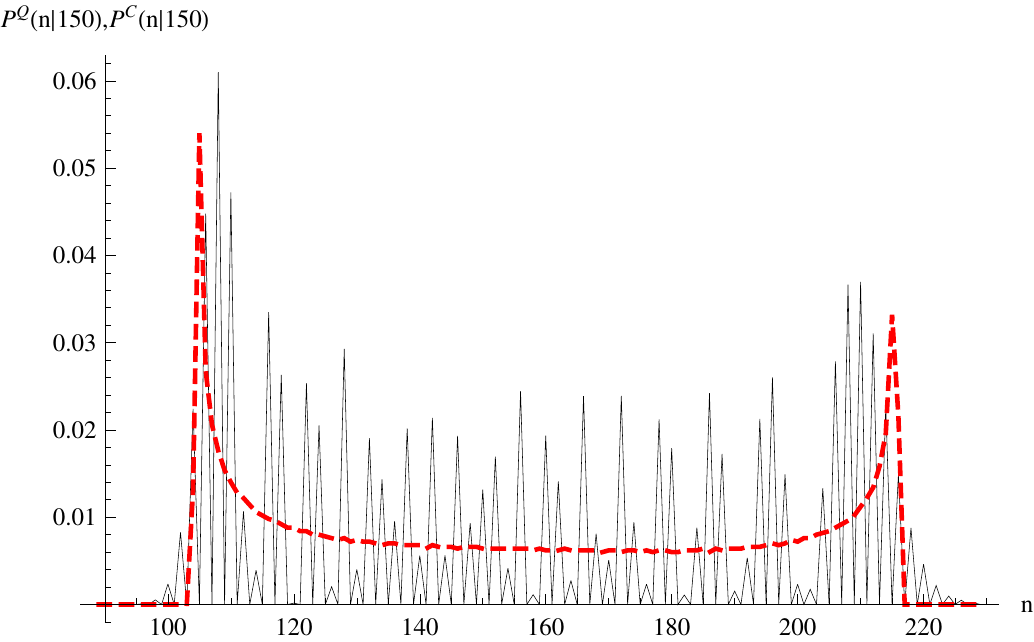}
\caption{Quantum (\ref{eq:Pnm}) and classical (\ref{eq:pcnm}) transition probabilities for the forced quartic oscillator.  The solid black curve shows $P^Q(n|m)$, while the dashed red  curve shows $P^C(n|m)$, for $m=150$.}
\label{comparison}
\end{figure}

To evaluate $P^C(n \vert m)$ we simulated $10^4$ Hamiltonian trajectories evolving under $H(z;\lambda_t)$.
Initial conditions were sampled microcanonically, and final conditions were binned according to the energy intervals of $H_B(z)$, as defined in Sec. \ref{subsec:csetup}.
For our initial microcanonical ensemble, we chose $m=150$, corresponding to $E_0=E_m^A = 1749.23$.
The results are shown by the dashed line in Fig.~\ref{comparison}. The transition probability is a smooth function of $n$ from $n_{\rm min} = 105$ to $n_{\rm max}=215$, and zero outside this range.

These sharp cutoffs reflect the fact that our initial conditions were sampled from a microcanonical distribution: $n_{\rm min}$ and $n_{\rm max}$ correspond to the minimal and maximal final energies that can be reached by trajectories launched with initial energies $E_0=E_m^A$, as illustrated in Fig.~\ref{fig:tangencies} of Sec.~\ref{sec:airy}.
The characteristic U-shape of the classical work distribution within the allowed region is related to the fact that the curves ${\cal A}$ and ${\cal B}$ shown in Fig.~\ref{Atau_B} become tangent to one another at $E=E_{\rm min}$ and $E=E_{\rm max}$ (again, see Fig.~\ref{fig:tangencies}).
For an exactly solvable model in which the same U-shape appears, see Fig.~1 of Ref.~\cite{Campisi2008b}.

For the quantum system, we expand a wavefunction evolving under $\hat{H}(\lambda_t)=\hat{p}^{2}/2M+\lambda_t \hat{q}^{4}$ as follows:
\begin{equation}
\psi(q,t)=\sum_{n}c_{n}(t) \, \langle q \vert \phi_{n}(\lambda_t) \rangle \, e^{-i \gamma_n(t)}
\qquad,\qquad
\gamma_n(t) = \frac{1}{\hbar} \int_{0}^{t}E_{n}(\lambda_{t^\prime}) {\rm d}t^\prime \quad.
\end{equation}
Here $\hat H \vert \phi_n \rangle = E_{n} \vert \phi_n \rangle$ and the $c_{n}$'s are expansion coefficients.
The time-dependent Schr\"odinger equation $i \hbar \dot{\psi}=\hat{H}\psi$ then produces the set of coupled ordinary differential equations,
\begin{equation}
\dot{c}_{n} = -  \dot{\lambda} \sum_{k}\left\langle \phi_{n} \left\vert \frac{\partial \phi_{k}}{\partial \lambda} \right. \right\rangle e^{i(\gamma_n-\gamma_k)} c_{k}
= -  \dot{\lambda} \sum_{k\ne n}  \frac{ \left\langle \phi_n \left\vert q^4 \right\vert \phi_k \right\rangle }{E_{k}-E_{n}}  e^{i(\gamma_n-\gamma_k)} c_{k} \quad.
\label{TDSE}
\end{equation}
We integrated these equations numerically, from initial conditions $c_n(0) = \delta_{mn}$, to obtain
\begin{equation}
P^Q(n \vert m) = \vert c_{n}(\tau) \vert^{2}.
\label{quantumprobability}
\end{equation}

In Fig.~\ref{comparison} we plot the quantum transition probability as a function of the final quantum number $n$ (solid line).
Although a correspondence between $P^Q(n \vert m)$ and $P^C(n \vert m)$ is visually evident, the quantum and classical cases differ in two distinct ways: (i) the quantum probability oscillates rapidly with $n$, and (ii) $P^Q(n \vert m)$ shows tails that ``tunnel'' into the classically forbidden regions $n<n_{\rm min}$ and $n>n_{\rm max}$.
Both features trace their origin to the wave nature of the quantum system.
We also see in Fig.~\ref{comparison} that $P^Q(n \vert m)=0$ when $n-m$ is an odd.
This result follows from the fact that $\left\langle \phi_n \left\vert q^4 \right\vert \phi_k \right\rangle = 0$ when the parities of $n$ and $k$ differ, hence only transitions between states of the same parity occur under Eq.~\ref{TDSE}.

\begin{figure}[tbh]
\includegraphics[trim = 0in 0in 0in 0in , scale=1.0,angle=0]{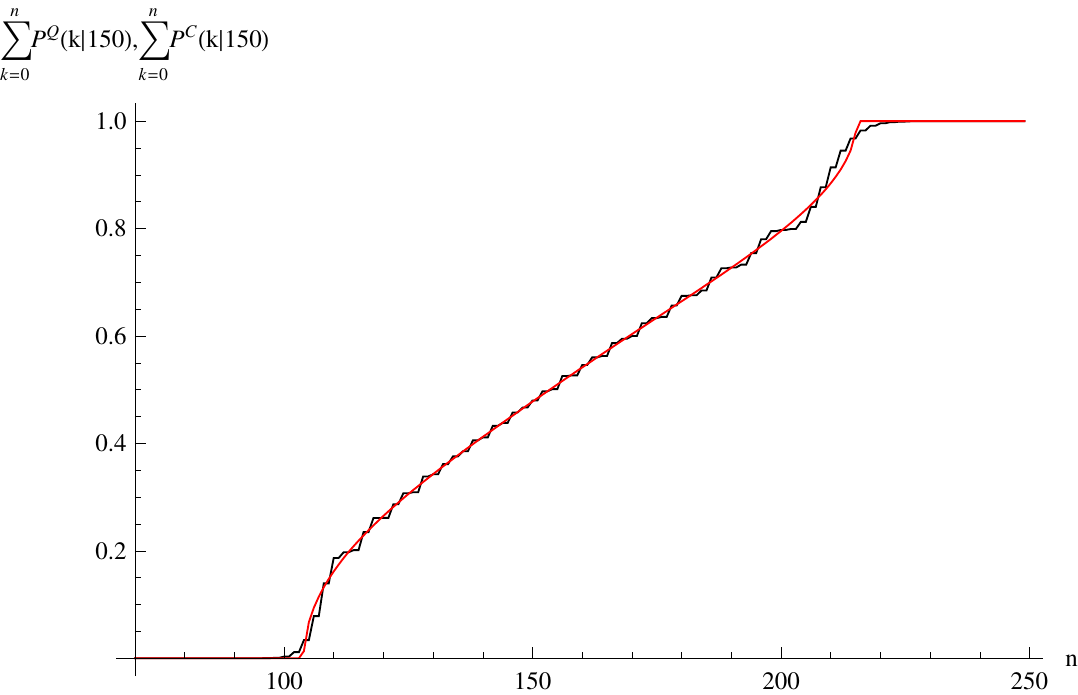}
\caption{Accumulated transition probabilities for the forced quartic oscillator. The jagged black curve represents the quantum case, $\sum_{k=0}^{n} P^Q(k \vert m)$, while the smooth red curve represents the classical case, $\sum_{k=0}^{n} P^C(k \vert m)$.}
\label{transitionprobabilities}
\end{figure}

In Fig.~\ref{transitionprobabilities} we plot the {\it accumulated} transition probabilities $\sum_{k=0}^{n} P^Q(k \vert m)$ and $\sum_{k=0}^{n} P^C(k \vert m)$, thereby smoothing out the rapid oscillations in the quantum transition probability. 
The close agreement observed in Fig.~\ref{transitionprobabilities} suggests that Eq.~\ref{eq:qwork} is indeed the appropriate quantum counterpart of Eq.~\ref{eq:cwork}, though distinctly non-classical features are visible in the interference and tunneling effects in Fig.~\ref{comparison}.

In Sections \ref{sec:semiclassical} and \ref{sec:airy} we investigate these issues analytically.
We will find that both the agreement seen in Fig.~\ref{transitionprobabilities} and the quantum effects evident in Fig.~\ref{comparison} can be understood quantitatively, within a semiclassical interpretation in which quantum dynamics are approximated by classical trajectories bearing time-dependent phases.

\section{Semiclassical theory}
\label{sec:semiclassical}

In Sec. \ref{subsec:ctp} we rewrite the classical transition probability $P^C(n|m)$ (Eq.~\ref{eq:pcnm}) as a sum over trajectories that begin and end with energies $E_m^A$ and $E_n^B$, respectively (Eq.~\ref{classical}).
In Sec. \ref{subsec:sctp} we use time-dependent WKB theory to derive a semiclassical approximation $P^{SC}(n\vert m)$ for the quantum transition probability $P^Q(n\vert m)$.
Our main result, Eq.~\ref{sc}, is expressed as a sum over the same trajectories that contribute to $P^C(n|m)$, only now each of these trajectories carries a phase, resulting in interference effects between different trajectories.
When these interferences are ignored, we recover the classical transition probability (Eq.~\ref{eq:Pnm_result}).
When they are included, the interferences give rise to the oscillations observed in the quantum transition probability.
In particular, interferences between symmetry-related trajectories account for the fastest oscillations observed in Fig.~\ref{comparison}, as we discuss in Sec.~\ref{subsec:interferences}.

\subsection{Classical transition probabilities}
\label{subsec:ctp}

We begin with the conditional probability distribution $\bar P^C(E\vert E_0)$ (Eq.~\ref{eq:PEE0}), where we have dropped the subscript $\tau$ from the final energy, for convenience.
As discussed in Sec.~\ref{sec:setup}, we imagine a swarm of trajectories evolving in time under $H(z;\lambda_t)$, with initial conditions sampled from the microcanonical phase space distribution
\begin{equation}
\eta_A(z;E_0) = \frac{ \delta\left[ E_0 - H_A(z) \right] } { \int {\rm d}z \, \delta\left[ E_0 - H_A(z) \right] }  = \frac{\delta\left[ E_0 - H_A(z) \right]}{h g_A(E_0)}  \quad ,
\end{equation}
where $E_0$ is a parameter of the distribution.
The quantity $\bar P^C(E\vert E_0) \, {\rm d}E$ is the fraction of these trajectories that end with a final energy in the infintesimal range $(E,E+{\rm d}E)$.

As in Fig.~\ref{fig:LagrangianManifolds}, let ${\cal A}_t$ denote the time-dependent curve that evolves from the initial energy shell $H_A = E_0$, and let ${\cal B}$ denote an energy shell of the final Hamiltonian, $H_B=E$.
The evolving curve ${\cal A}_t$ can be described by a multivalued, time-dependent momentum field $p_b^{{\cal A}_t}(q)$, where the index $b$ labels the branches of ${\cal A}_t$ at the coordinate value $q$.
E.g.\  in Fig.~\ref{Atau_B} the momentum field for ${\cal A}_\tau$ has two branches at $q=q_1$.
Similarly, the multivalued momentum field $p_b^{\cal B}(q)$ describes the fixed energy shell $H_B=E$, whose two branches are
\begin{equation}
\label{eq:branches}
p_b^{\cal B}(q)=\pm\sqrt{2M[E-V(q;B)]} \qquad (b=\pm).
\end{equation}
For convenience, we will henceforth use the notation ${\cal A}$ (without a subscript) to indicate the surface ${\cal A}_\tau$.
The points of intersection between ${\cal A}$ and ${\cal B}$, highlighted by dots in Fig.~\ref{Atau_B}, represent the trajectories that contribute to $\bar P^C(E\vert E_0)$.

Our swarm of trajectories at time $t$ is described not only by the evolving surface ${\cal A}_t$, but also by a probability density on that surface.
If we project this density onto the $q$-axis, the projected density $\rho^{{\cal A}_t}(q)$ is a sum over the branches of the surface ${\cal A}_t$:
\begin{equation}
\rho^{{\cal A}_t}(q) = \sum_b \rho_b^{{\cal A}_t}(q).
\end{equation}
The fixed microcanonical ensemble corresponding to $H_B=E$ can similarly be projected onto a density
\begin{equation}
\label{eq:rhoBdensity}
\rho^{\cal B}(q) 
= \frac{1}{h g_B(E)} \int {\rm d}p \, \delta\left[ E - H_B(q,p) \right]
= \frac{1}{h g_B(E)} \sum_b \left\vert \frac{\partial H_B}{\partial p} \right\vert_b^{-1} \equiv \sum_b \rho_b^{\cal B}(q)
\end{equation}
with $\partial H_B/\partial p = p/M$ evaluated at $p = p_b^{\cal B}(q)$.

Let $l = 1, 2, \cdots K$ be an index labeling the intersection points of ${\cal A}$ and ${\cal B}$ (e.g. $K=4$ in Fig.~\ref{fig:LagrangianManifolds}), and let $b_l$ denote the branch corresponding to the $l'th$ intersection point:
\begin{equation}
\label{eq:qlpl}
(q_l,p_l) = \left( q_l, p_{b_l}^{{\cal A}}(q_l) \right) = \left( q_l, p_{b_l}^{\cal B}(q_l) \right).
\end{equation}
With some abuse of notation, we have written $b_l$ rather than $b_l^{{\cal A}}$ and $b_l^{\cal B}$, though in general the $l$'th intersection point may occur at differently indexed branches of the two surfaces.

\begin{figure}[tbp]
\includegraphics[trim = 1in 0in 0in 0in , scale=0.3,angle=0]{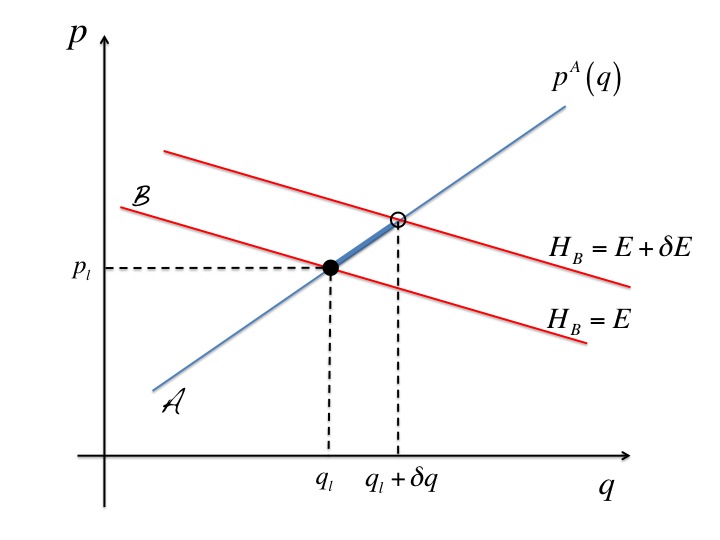}
\caption{The filled circle indicates an intersection between the closed curves ${\cal A}$ and ${\cal B}$, see Fig.~\ref{fig:LagrangianManifolds}.  The open circle is an intersection point between ${\cal A}$ and a nearby energy shell $H_B = E+{\delta E}$.}
\label{fig:intersection}
\end{figure}

To evaluate $\bar P^C(E\vert E_0)$, we begin with the contribution $\bar P_l^C(E\vert E_0)$ from a single intersection point, $l$.
Fig.~\ref{fig:intersection} depicts this intersection, together with an intersection point involving an infinitesimally displaced energy shell $H_B=E+\delta E$.
The highlighted line segment connecting these two points represents a set of trajectories with final energies between $E$ and $E+\delta E$, hence
\begin{equation}
\label{eq:intersection}
\rho^{{\cal A}}(q_l) \cdot \vert \delta q \vert = \bar P_l^C(E\vert E_0) \cdot \vert \delta E\vert.
\end{equation}
Since we focus here on the contribution from a single intersection between ${\cal A}$ and ${\cal B}$, we suppress the subscripts $b_l$ indicating the branches of these surfaces.
By construction, the ratio $\delta E/\delta q$ is the rate of change of $H_B$ with respect to $q$ along the curve ${\cal A}$:
\begin{equation}
\label{eq:dEdq}
\frac{\delta E}{\delta q} = \frac{\rm d}{{\rm d}q} H_B\left(q,p^{{\cal A}}(q)\right) \Bigr\vert_{q=q_l}.
\end{equation}
Since $p^{\cal B}(q)$ specifies a surface of constant $H_B$, we have
\begin{equation}
\frac{\partial p^{\cal B}}{\partial q} = \frac{\partial p}{\partial q} \Bigr\vert_{H_B} = - \frac{\partial H_B/\partial q}{\partial H_B/\partial p}
\end{equation}
hence
\begin{equation}
\label{eq:dHBdq}
\frac{\rm d}{{\rm d}q} H_B\left(q,p^{{\cal A}}(q)\right)
= \frac{\partial H_B}{\partial q} + \frac{\partial H_B}{\partial p} \frac{\partial p^{{\cal A}}}{\partial q}
= \frac{\partial H_B}{\partial p} \left( \frac{\partial p^{{\cal A}}}{\partial q} - \frac{\partial p^{\cal B}}{\partial q} \right).
\end{equation}
Combining Eqs.~(\ref{eq:rhoBdensity}), (\ref{eq:intersection}), (\ref{eq:dEdq}), and (\ref{eq:dHBdq}) we get
\begin{equation}
\bar P_l^C(E\vert E_0) =
\rho^{{\cal A}}(q_l) \cdot \left\vert \frac{\delta E}{\delta q} \right\vert^{-1} =
h g_B(E) \, \rho^{{\cal A}}(q_l) \rho^{\cal B}(q_l) \, \left\vert \frac{\partial p^{\cal A}}{\partial q} - \frac{\partial p^{\cal B}}{\partial q} \right\vert_{q=q_l}^{-1}.
\end{equation}
Summing over all intersection points gives us
\begin{equation}
\label{eq:finalResult_classical}
\bar P^C(E\vert E_0) = h g_B(E) \, \sum_{l=1}^K
\rho_{b_l}^{{\cal A}}(q_l) \rho_{b_l}^{\cal B}(q_l) \, \left\vert \frac{\partial p_{b_l}^{{\cal A}}}{\partial q} - \frac{\partial p_{b_l}^{\cal B}}{\partial q} \right\vert_{q=q_l}^{-1} \quad.
\end{equation}
If we now set $E_0 = E_m^A$ and $E = E_n^B$, then Eq.~\ref{eq:pcnm} gives us
\begin{equation}
P^C(n|m)= h \sum_{l=1}^K \rho^{\cal A}_{b_{l}}(q_l) \rho^{\cal B}_{b_{l}}(q_l)  \left| \frac{\partial p^{\cal A}_{b_{l}} }{\partial q} - \frac{\partial p^{\cal B}_{b_{l}} }{\partial q} \right|_{q=q_{l}}^{-1}.
\label{classical}
\end{equation}
This is the main result of this subsection.

We will use the term {\it classically allowed region} to denote the range $E_{\rm min} \le E \le E_{\rm max}$ within which the function $\bar P^C(E\vert E_0)$ does not vanish, representing all final energies that can be attained by trajectories with initial energy $E_0$.
In the coarse-grained function $P^C(n\vert m)$, the classically allowed region is bracketed by the values $n_{\rm min}$ and $n_{\rm max}$, denoting the energy intervals of $H_B$ containing $E_{\rm min}$ and $E_{\rm max}$.
As illustrated in Fig.~\ref{fig:tangencies} below, the values $E_{\rm min}$ and $E_{\rm max}$ correspond, respectively, to the largest energy shell of $H_B$ that fits entirely within the closed curve ${\cal A}$, and the smallest energy shell of $H_B$ that surrounds the entire curve ${\cal A}$.
At these two energies, the curves ${\cal A}$ and ${\cal B}$ become tangent to one another, $\partial p_{b_l}^{\cal A}/\partial q = \partial p_{b_l}^{\cal B}/\partial q$, leading to divergences in Eq.~(\ref{eq:finalResult_classical}), which are reflected in the two sharp peaks in $P^C(n|m)$ in Fig. \ref{comparison}.  Similar divergences appear in a more familiar context, namely the density $\rho^{\cal B}(q)$ that describes the projection of a fixed microcanonical ensemble $H_B=E$ onto the coordinate axis (see Eq.~(\ref{eq:rhoBdensity})).
In this case the divergences occur at classical turning points at which the velocity $\partial H_B/\partial p$ vanishes.

In the example introduced in Sec.~\ref{sec:numerical}, the reflection symmetry of the quartic potential implies that trajectories come in symmetry-related pairs: if $(q_t,p_t)$ is a solution of Hamilton's equations under the time-dependent Hamiltonian $H(q,p;\lambda_t)$, then so is $(-q_t,-p_t)$.
This means that for every intersection point $(q_l,p_l)$ between ${\cal A}$ and ${\cal B}$, there will be another intersection point at $(-q_l,-p_l)$, and the two will contribute equally to $P^C(n|m)$.

\subsection{Semiclassical transition probabilities}
\label{subsec:sctp}

Let us now evaluate $\bar P^Q(n\vert m)$ in the semiclassical limit.
Appendix \ref{A} provides a brief introduction to time-dependent WKB theory, and for further details we refer to the reviews by Delos~\cite{Delos1986} and Littlejohn~\cite{Littlejohn1992}.
Using Eq. \ref{eq:WKBsoln}, the wavefunction $\psi(q,t) = \langle q \vert \hat U_t \vert \phi_m^A \rangle$ and the $n$'th eigenstate of $\hat H_B$ can be written as
\begin{subequations}
\label{eq:wkbs}
\begin{eqnarray}
\psi(q,t) &=& \sum_b \sqrt{\rho_b^{{\cal A}_t}(q)} \exp \left[ \frac{i}{\hbar} S_b^{{\cal A}_t}(q) - i \mu_b^{{\cal A}_t} \frac{\pi}2 \right], \\
\label{eq:eigenstate}
\phi_n^B(q) &=& \sum_b \sqrt{\rho_b^{\cal B}(q)} \exp \left[ \frac{i}{\hbar} S_b^{\cal B}(q) - i \mu_b^{\cal B} \frac{\pi}2 \right],
\end{eqnarray}
\end{subequations}
where the actions $S_b^{{\cal A}_t}(q)$ and $S_b^{\cal B}(q)$ generate the momentum fields
\begin{equation}
\label{eq:p=dSdq}
p_b^{{\cal A}_t}(q) = \frac{\partial S_b^{{\cal A}_t}}{\partial q}
\quad,\quad
p_b^{\cal B}(q) = \frac{\partial S_b^{\cal B}}{\partial q},
\end{equation}
and the integers $\mu_b^{{\cal A}_t}$ and $\mu_b^{\cal B}$ are {\it Maslov indices} (see Appendix \ref{A}).
The densities and momentum fields in Eqs.~\ref{eq:wkbs} and \ref{eq:p=dSdq} are the same as those appearing in Sec. \ref{subsec:ctp}.

From Eq.~(\ref{eq:wkbs}) we get
\begin{equation}
\label{eq:innerProduct}
\left\langle \phi_n^B \vert  \psi_\tau \right\rangle = \sum_{b,b^\prime} \, \int {\rm d}q \,
\sqrt{\rho_b^{\cal A} \, \rho_{b^\prime}^{\cal B} } \, \exp \left[ \frac{i}{\hbar} \left( S_b^{\cal A} - S_{b^\prime}^{\cal B} \right)
-i \left( \mu_b^{\cal A} - \mu_{b^\prime}^{\cal B} \right) \frac{\pi}2 \right] \quad,
\end{equation}
which appears as Eq.~37b in Ref.~\cite{Schwieters1995b}.
Using the stationary phase approximation to evaluate the integral, we obtain a sum of contributions from points $q_l$ satisfying the condition
\begin{equation}
\label{eq:SPcondition}
\frac{\partial S_b^{\cal A}}{\partial q}(q_l) = \frac{\partial S_{b^\prime}^{\cal B}}{\partial q}(q_l)
\end{equation}
which is equivalent to
\begin{equation}
p_b^{\cal A}(q_l) = p_{b^\prime}^{\cal B}(q_l) \equiv p_l \quad.
\end{equation}
Hence $\langle \phi_n^B \vert  \psi_\tau \rangle$ reduces to a sum of contributions arising from the $K$ intersection points of the surfaces ${\cal A}$ and ${\cal B}$, representing trajectories with initial and final energies $E_m^A$ and $E_n^B$, just as in the classical case (Eq.~\ref{eq:qlpl}).

To determine the contribution from the $l$'th intersection point, we perform a quadratic expansion around $q=q_l$:
\begin{equation}
S_{b_l}^{\cal A}(q) - S_{b_l}^{\cal B}(q) \approx \Delta S_l + \frac{1}{2}\kappa_l (q-q_l)^2 \quad,
\label{quadratic}
\end{equation}
with $\Delta S_l = S_{b_l}^{\cal A}(q_l) - S_{b_l}^{\cal B}(q_l)$ and
\begin{equation}
\kappa_l = \frac{\partial^2 S_{b_l}^{\cal A}}{\partial q^2}(q_l) - \frac{\partial^2 S_{b_l}^{\cal B}}{\partial q^2}(q_l)
= \frac{\partial p_{b_l}^{\cal A}}{\partial q}(q_l) - \frac{\partial p_{b_l}^{\cal B}}{\partial q}(q_l).
\end{equation}
Performing the stationary phase integral then gives us:
\begin{equation}
\label{eq:stationaryPhaseIntegral}
\begin{split}
\left\langle \phi_n^B \vert  \psi_\tau \right\rangle
&\approx
\sum_{l=1}^K
\sqrt{\rho_{b_l}^{\cal A} \, \rho_{b_l}^{\cal B} } \,
\exp \left( \frac{i}{\hbar} \Delta S_l  -i\Delta\mu_l \frac{\pi}{2} \right)
\int {\rm d}q \, \exp \left[ \frac{i\kappa_l}{2\hbar} (q-q_l)^2 \right] \\
&= \sum_{l=1}^K
\sqrt{\rho_{b_l}^{\cal A} \, \rho_{b_l}^{\cal B} } \,
\exp \left( \frac{i}{\hbar} \Delta S_l  -i\Delta\mu_l \frac{\pi}{2} \right)
\sqrt{\frac{2\pi\hbar}{\vert\kappa_l\vert}} e^{i\sigma_l\pi/4}
= \sum_l  a_l e^{i\theta_l}
\quad ,
\end{split}
\end{equation}
where $\rho_{b_l}^{\cal A,\cal B}$ are evaluated at $q=q_l$; $\Delta\mu_l = \mu_{b_l}^{\cal A} - \mu_{b_l}^{\cal B}$; and $\sigma_l = {\rm sign}(\kappa_l) = \pm 1$.
In Ref.~\cite{Schwieters1995b}, where similar calculations are performed, a quantity equivalent to our $\pi\sigma_l/4$ is identified as a Maslov-like phase (Ref.~\cite{Schwieters1995b}, p. 1037).

Eq.~(\ref{eq:stationaryPhaseIntegral}) expresses $\langle \phi_n^B \vert  \psi_\tau \rangle$ as a sum over trajectories, each contributing an amplitude and a phase:
\begin{equation}
\label{eq:thetadef}
a_l = \sqrt{  2\pi\hbar \, \frac{  \rho_{b_l}^{\cal A} \, \rho_{b_l}^{\cal B}  } { \vert \kappa_l \vert  }  } \quad , \quad
\theta_l = \frac{1}{\hbar} \Delta S_l - \frac{\pi}{2} \Delta\mu_l  + \frac{\pi}{4} \sigma_l \quad .
\end{equation}
The semiclassical transition probability $P^{SC}(n|m)\approx\left | \left \langle \phi_{n}^{B} \right | \left. \psi_\tau\right \rangle \right |^{2}$ is then given by
\begin{equation}
\label{sc}
P^{SC}(n|m)= \left\vert
\sum_{l=1}^K  a_l e^{i\theta_l}
 \right \vert^{2} \quad.
\end{equation}
The general form of Eq.~\ref{sc} is not surprising.
The path integral formulation of quantum mechanics tells us that solutions of the time-dependent Schr\" odinger equation can be represented in terms of sums over classical paths decorated by phases $\exp(iS/\hbar)$, and when $\hbar\rightarrow 0$ the dominant contributions to the sums come from those paths that are solutions of the classical equations of motion~\cite{FeynmanHibbs_book}.

\begin{figure}[tbh]
\includegraphics[trim = 0in 0in 0in 0in , scale=1.0,angle=0]{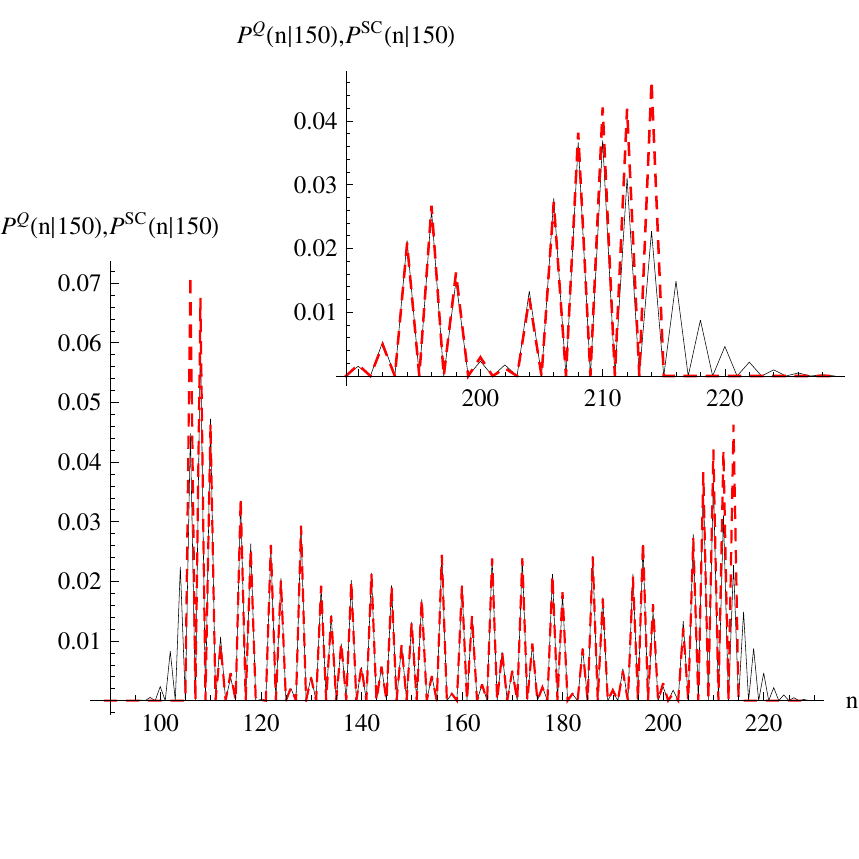}
\caption{A comparison of quantum and semiclassical transition probabilities for our model system. The black solid curve shows $P^Q(n|m)$, while the red dashed curve shows $P^{SC}(n|m)$, as given by Eq.~\ref{sc}. All the parameters are the same as those in Fig.~\ref{comparison}.}
\label{sctransitionprobabilities}
\end{figure}

We expect the semiclassical transition probability function $P^{SC}(n|m)$ to provide a good approximation to the quantum transition probability $P^Q(n|m)$, for large quantum numbers.
For the forced quartic oscillator, Fig.~\ref{sctransitionprobabilities} compares these functions, taking $m=150$.
$P^Q$ was computed as described in Sec.~\ref{sec:numerical}, and $P^{SC}$ was evaluated directly from classical simulations of trajectories evolving from an initial microcanonical ensemble.
(The evolution of the multivalued action $S^{{\cal A}_t}$ was obtained by integrating $p\,{\rm d}q - H\,{\rm d}t$ along each trajectory, see e.g.\ Sec. 3.5 of Ref.~\cite{Littlejohn1992}.)
We observe that the agreement is excellent, except in the vicinity of the boundaries between the classically allowed and forbidden regions.
We will examine these boundaries in more detail in Sec. \ref{sec:airy}.

Now let us simplify Eq.~\ref{sc} by simply ignoring the cross terms in the double sum.
This is the so-called {\it diagonal approximation} \cite{Berry1985,Doron1991,Baranger1993}, and it leads to
\begin{equation}
\label{eq:Pnm_result}
P^{SC}(n|m) \stackrel{\rm diag}{\approx}  \sum_{l=1}^K a_l^2 = 2\pi\hbar \, \sum_{l=1}^K \rho_{b_l}^{\cal A} \, \rho_{b_l}^{\cal B} \,
\left\vert \frac{\partial p_{b_l}^{\cal A}}{\partial q} - \frac{\partial p_{b_l}^{\cal B}}{\partial q} \right\vert_{q=q_l}^{-1} \quad,
\end{equation}
which is identical to the expression for the classical transition probability, $P^C(n|m)$ (Eq.~\ref{classical}).

The agreement between $P^Q$ and $P^{SC} = \vert \sum a_l e^{i\theta_l} \vert^2$ seen in Fig.~\ref{sctransitionprobabilities} suggests that the rapid oscillations of the quantum probability distribution can be understood in terms of phase interference between different trajectories.
When these interferences are ignored, we recover the classical probability distribution, as we see in Eq.~\ref{eq:Pnm_result}.
Hence $P^{SC}$ acts as a bridge to between the quantum and classical transition probabilities,
\begin{equation}
P^Q(n|m)  \approx P^{SC}(n|m) \stackrel{\rm diag}{\approx}    P^C(n|m) \quad ,
\label{convergence}
\end{equation}
and thus between the quantum and classical work distributions (Eqs.~\ref{eq:qworkDist}, \ref{eq:cworkDist_sum}).
The first approximation in Eq.~\ref{convergence} involves time-dependent WKB theory together with a stationary phase evaluation of integrals, and the second is the diagonal approximation in which interferences are ignored.

\subsection{Interferences and symmetries}
\label{subsec:interferences}

Let us now examine the interferences in more detail.
Recall from Sec.~\ref{subsec:ctp} that the trajectories contributing to Eq.~\ref{sc} come in symmetry-related pairs.
This allows us to write:
\begin{equation}
\label{eq:symmetrySum}
\left\langle \phi_n^B \vert  \psi_\tau \right\rangle =
\sum_{l=1}^K a_l e^{i\theta_l} = \sum_{l=1}^J \left( a_l e^{i\theta_l} + a_{l+J} e^{i\theta_{l+J}} \right) 
\end{equation}
where $J=K/2$ is an integer, and the $l$'th and $(l+J)$'th trajectories are related by symmetry:
\begin{equation}
\label{eq:mirror}
(q_{l+J},p_{l+J}) = (-q_l,-p_l) \quad.
\end{equation}
Here we have indexed the points $(q_l,p_l)$ from $l=1$ to $K$, according to the order in which they appear as we proceed clockwise around the energy shell ${\cal B}$, first rightward along the upper branch and then leftward along the lower branch. 
Eq.~\ref{eq:mirror} implies
\begin{equation}
\label{eq:sigmaSym}
a_{l+J} = a_l \quad,\quad
\sigma_{l+J} = \sigma_l \quad.
\end{equation}
Using the fact that the points $l$ and $l+J$ are located directly opposite one another on the closed curves ${\cal A}$ and ${\cal B}$ (Eq.~\ref{eq:mirror}), we obtain
\begin{equation}
\label{eq:deltaS}
S_{l+J}^{\cal A}-S_l^{\cal A} = \frac{1}{2} \oint_{\cal A} p \, {\rm d}q =  \left( m+\frac{1}{2} \right) \pi\hbar \quad,\quad
S_{l+J}^{\cal B}-S_l^{\cal B} = \frac{1}{2} \oint_{\cal B} p \, {\rm d}q = \left( n+\frac{1}{2} \right) \pi\hbar
\end{equation}
and
\begin{equation}
\label{eq:deltamu}
\mu_{l+J}^{\cal A} - \mu_{l}^{\cal A} = \mu_{l+J}^{\cal B} - \mu_{l}^{\cal B} = 1 \quad ,
\end{equation}
hence
\begin{equation}
\label{eq:SmuSym}
\Delta S_{l+J} = \Delta S_l + (m-n)\pi\hbar
\quad,\quad
\Delta\mu_{l+J} = \Delta\mu_l  \quad .
\end{equation}
(In Eq.~\ref{eq:deltaS} we have invoked the semiclassical quantization condition, Eq.~\ref{eq:midpoint}, together with Liouville's theorem, which guarantees that $\oint_{{\cal A}_t} p\,{\rm d}q$ remains constant with time.
In Eq.~\ref{eq:deltamu} we have used the fact that the Maslov index is incremented by $+2$ as one proceeds around the closed curve ${\cal A}$ or ${\cal B}$, see e.g.\ Fig.~1 of Ref.~\cite{Schwieters1995b}.)
Eqs.~\ref{eq:thetadef}, \ref{eq:symmetrySum}, \ref{eq:sigmaSym} and \ref{eq:SmuSym} give us
\begin{equation}
\label{eq:fastInterference}
\left\langle \phi_n^B \vert  \psi_\tau \right\rangle = \sum_{l=1}^J a_l e^{i\theta_l} \left[ 1 + e^{i(m-n)\pi} \right] \quad .
\end{equation}
When $m-n$ is even, the symmetry-related trajectories interfere constructively, but when $m$ and $n$ have opposite parities they interfere destructively and we get $P^{SC}(n\vert m)=0$.
This provides a semiclassical explanation for the observation, noted in Sec.~\ref{sec:numerical}, that the quantum transition probability vanishes when $m-n$ is odd:
the most rapid oscillations in $P^Q(n\vert m)$ arise from interference between symmetry-related trajectories.
The same interpretation was given by Miller to explain angular momentum selection rules in the scattering matrix describing an atom impinging on a homonuclear diatomic molecule (see Sec. III C of Ref.~\cite{Miller1974}).

\begin{figure}[tbh]
	\subfigure[\ ]{\label{fig:fast}
	\includegraphics[scale=0.75]{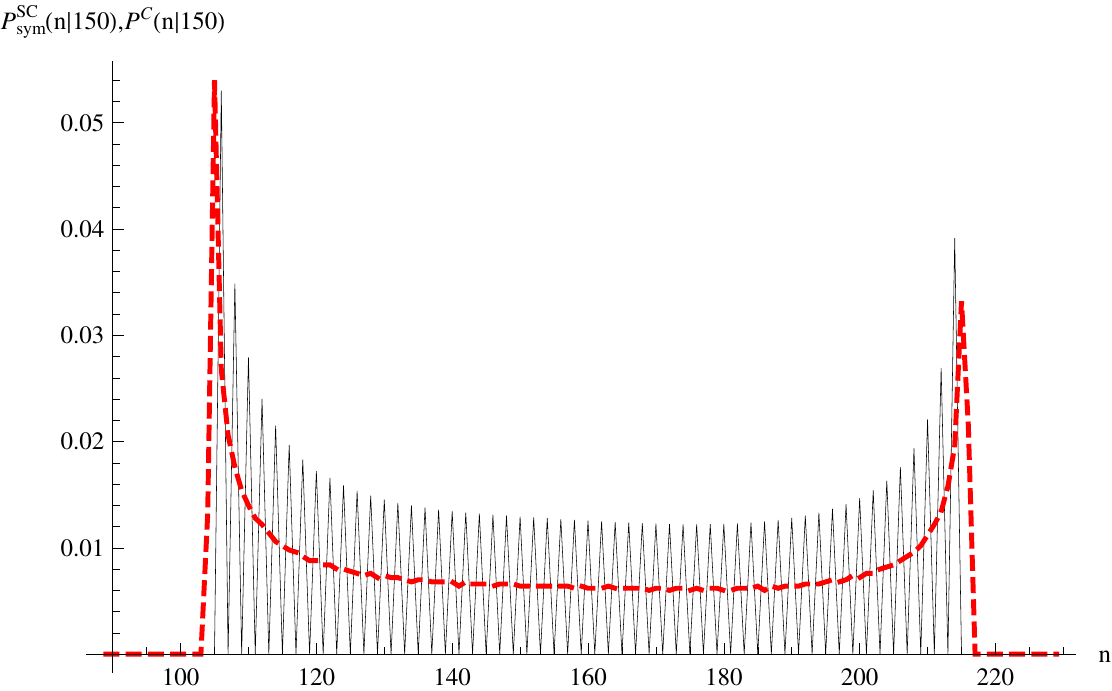}
	}
	\subfigure[\ ]{\label{fig:slow}
	\includegraphics[scale=0.75]{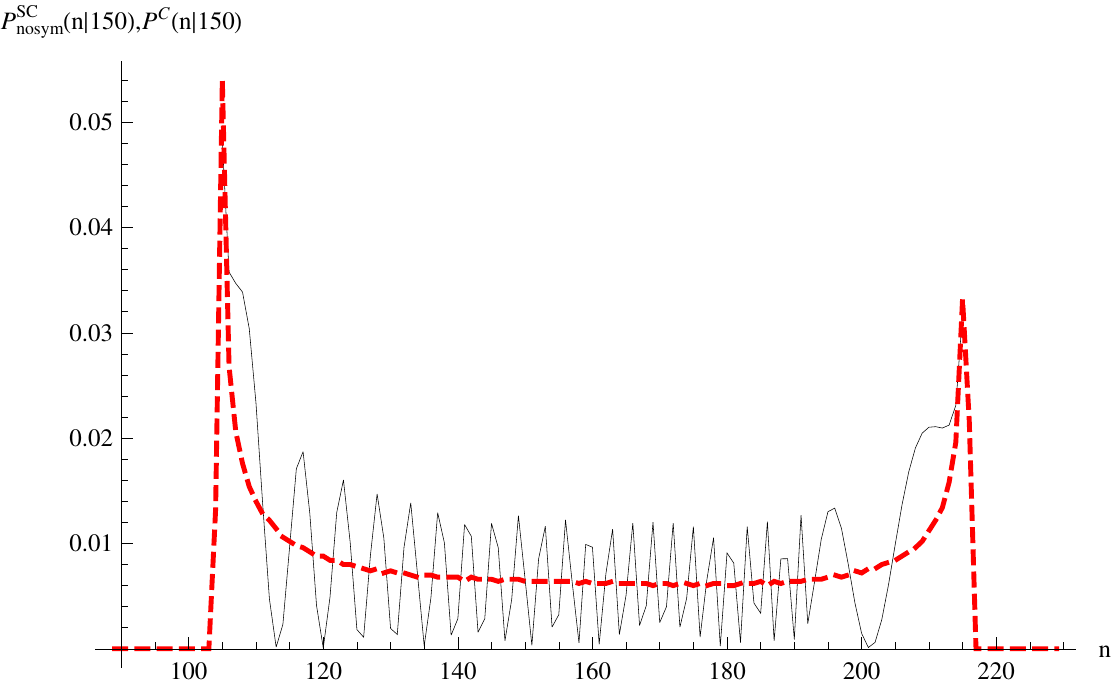}
	}
\caption{
Contributions to the interference pattern in $P^{SC}(n|m)$.
Figure (a) includes only interferences between symmetry-related trajectories, while (b) includes only interferences between non-symmetry-related trajectories.
}
\label{fig:partial_interference}
\end{figure}

The interference effects in $P^{SC}$, due to cross terms in Eq.~\ref{sc}, include contributions not only from pairs of trajectories that are related by symmetry (Eq.~\ref{eq:mirror}), but also from those that are not related by symmetry:
\begin{equation}
\label{eq:separateContributions}
P^{SC}(n|m) =  \sum_{l=1}^K a_l^2
+ {\sum_{l\ne k}}^{\rm s} a_l a_k e^{i(\theta_l-\theta_k)}
+ {\sum_{l\ne k}}^{\rm n} a_l a_k e^{i(\theta_l-\theta_k)}
\quad,
\end{equation}
where $\sum^{\rm s}$ denotes a sum of symmetry-related pairs of trajectories, $(q_k,p_k)=-(q_l,p_l)$, and $\sum^{\rm n}$ is a sum over non-symmetry-related pairs, $(q_k,p_k)\ne -(q_l,p_l)$.
To illustrate these contributions separately, we constructed the quantity $P_{\rm sym}^{SC}$ by omitting the sum $\sum^{\rm n}$ on the right side of Eq.~\ref{eq:separateContributions}, and we similarly constructed $P_{\rm nosym}^{SC}$ by omitting $\sum^{\rm s}$.
The results are shown in Fig.~\ref{fig:partial_interference}.
$P_{\rm sym}^{SC}$ displays a regular oscillation, as the symmetry-related trajectories are either exactly in phase or exactly out of phase, according to the parity of $m-n$.
The oscillations in $P_{\rm nosym}^{SC}$ are less regular, as the phases $\theta_l-\theta_k$ are not necessarily integer multiples of $\pi$.
When all cross terms are included, both sets of oscillations combine to give the pattern in Fig.~\ref{sctransitionprobabilities}.
For instance, the gap observed near $n=200$ in Fig.~\ref{sctransitionprobabilities} reflects the corresponding minimum in $P_{\rm nosym}^{SC}$ seen in Fig.~\ref{fig:partial_interference}(b).

\section{Airy tails}
\label{sec:airy}

In Fig.~\ref{sctransitionprobabilities} we see that that our semiclassical approximation fails around the boundaries of the classical allowed region at $n_{\rm min}$ and $n_{\rm max}$.
Although the semiclassical transition probability derived in Sec.~\ref{sec:semiclassical} vanishes identically outside the classically allowed region (since only classical trajectories contribute to $P^{SC}$), the quantum transition probability tunnels into the classically forbidden region, as seen in Fig.~\ref{sctransitionprobabilities}.
In this section we construct a semiclassical approximation that describes this behavior.
Our central results are given by Eqs.~\ref{eq:airy_single} and \ref{eq:airy_interference}, and illustrated in Fig.~\ref{perfect}.

Similar tunneling behavior is observed in the wavefunctions of energy eigenstates, which exhibit tails that reach into classically forbidden regions \cite{Griffiths1995}, as well as in semiclassical treatments of the $S$ matrix in molecular collisions \cite{Marcus1971}; in both cases the tails are approximated by Airy functions.
The expressions that we obtain in this section will also be expressed in terms of Airy functions.

\begin{figure}[tbh]
	\subfigure[\ ]{\label{fig:doubleWell_noSym}
	\includegraphics[scale=0.3]{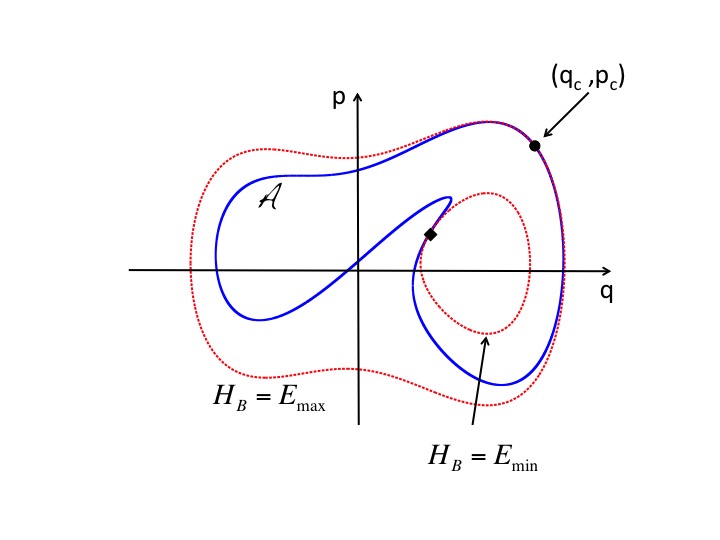}
	}
	\subfigure[\ ]{\label{fig:quartic_Sym}
	\includegraphics[scale=0.3]{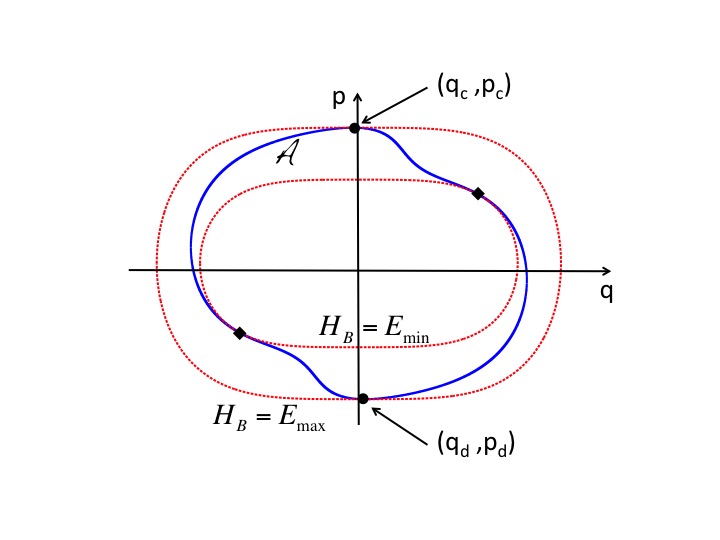}
	}
\caption{Solid blue lines depict the surfaces ${\cal A} = {\cal A}_\tau$ (as in Fig.~\ref{fig:LagrangianManifolds}).
Dashed red lines are energy shells ${\cal B}$ of $H_B$.
Diamonds and dots indicate the points of tangency between ${\cal A}$ and ${\cal B}$ at $E_{\rm min}$ and $E_{\rm max}$, respectively.
(a) In the absence of symmetries, we generically expect a single tangent point at $E_{\rm min}$ and $E_{\max}$.
(b) For the Hamiltonian that we studied numerically, Eq.~\ref{Hamiltonian}, the symmetry of the quartic potential implies that both ${\cal A}$ and the energy shells ${\cal B}$ are symmetric under $(q,p) \rightarrow (-q,-p)$, thus points of tangency come in symmetry-related pairs.
}
\label{fig:tangencies}
\end{figure}

The failure of $P^{SC}(n \vert m)$ in the boundary regions originates in the fact that the curves ${\cal A}$ and ${\cal B}$ become tangent to one another at $E=E_{\rm min}$ and $E=E_{\rm max}$.
In the example we have studied numerically the curves are tangent simultaneously at two points, due to the symmetry of the quartic potential, as shown in Fig.~\ref{fig:quartic_Sym}.
However, it is useful first to discuss the generic case, where there exists only a single point of tangency, depicted in Fig.~\ref{fig:doubleWell_noSym}.
For specificity we will discuss the upper boundary of the classically allowed region at $E_{\rm max}$, but similar comments apply to the lower boundary at $E_{\rm min}$.

\begin{figure}[tbh]
	\subfigure{\label{fig:Delta_p}
	\includegraphics[scale=0.3]{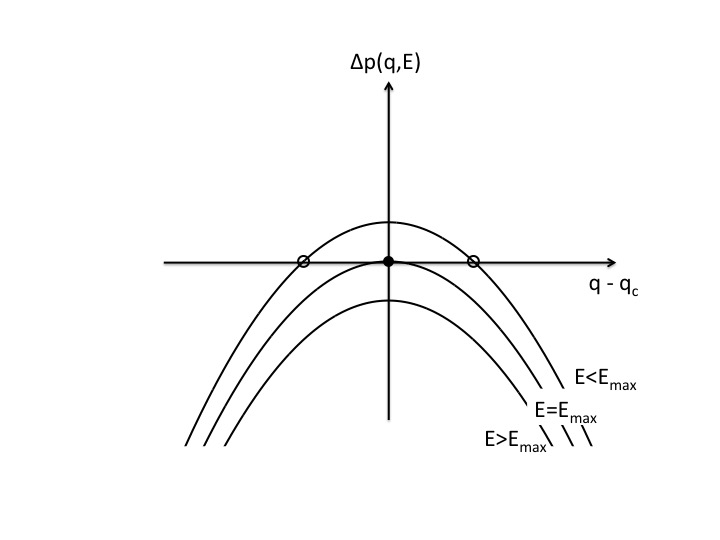}
	}
	\subfigure{\label{fig:cubic}
	\includegraphics[scale=0.3]{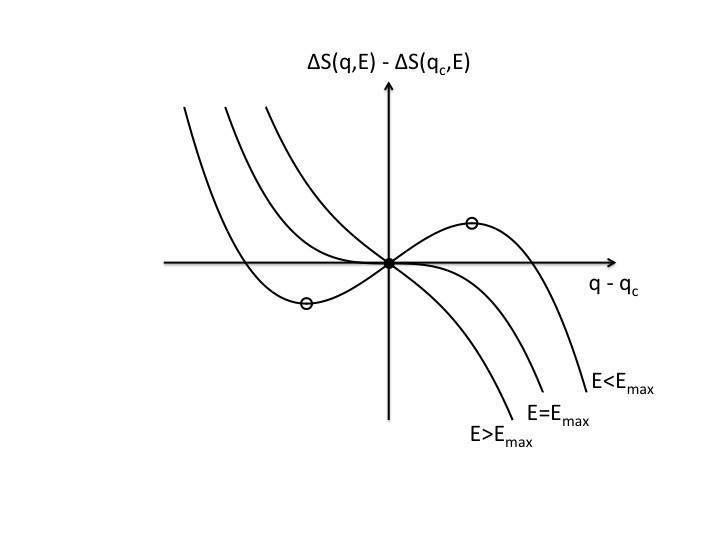}
	}
\caption{Behavior of $\Delta p$ and $\Delta S$ for $E \approx E_{\rm max}$.  The open circles correspond to intersection points between ${\cal A}$ and ${\cal B}$ when $E<E_{\rm max}$. These coalesce into a single point of tangency at $E=E_{\rm max}$ (solid circle), and for $E>E_{\rm max}$, ${\cal A}$ and ${\cal B}$ do not intersect.}
\label{fig:coalescence}
\end{figure}

\begin{figure}[tbh]
\includegraphics[trim = 0in 0in 0in 0in , scale=1.0,angle=0]{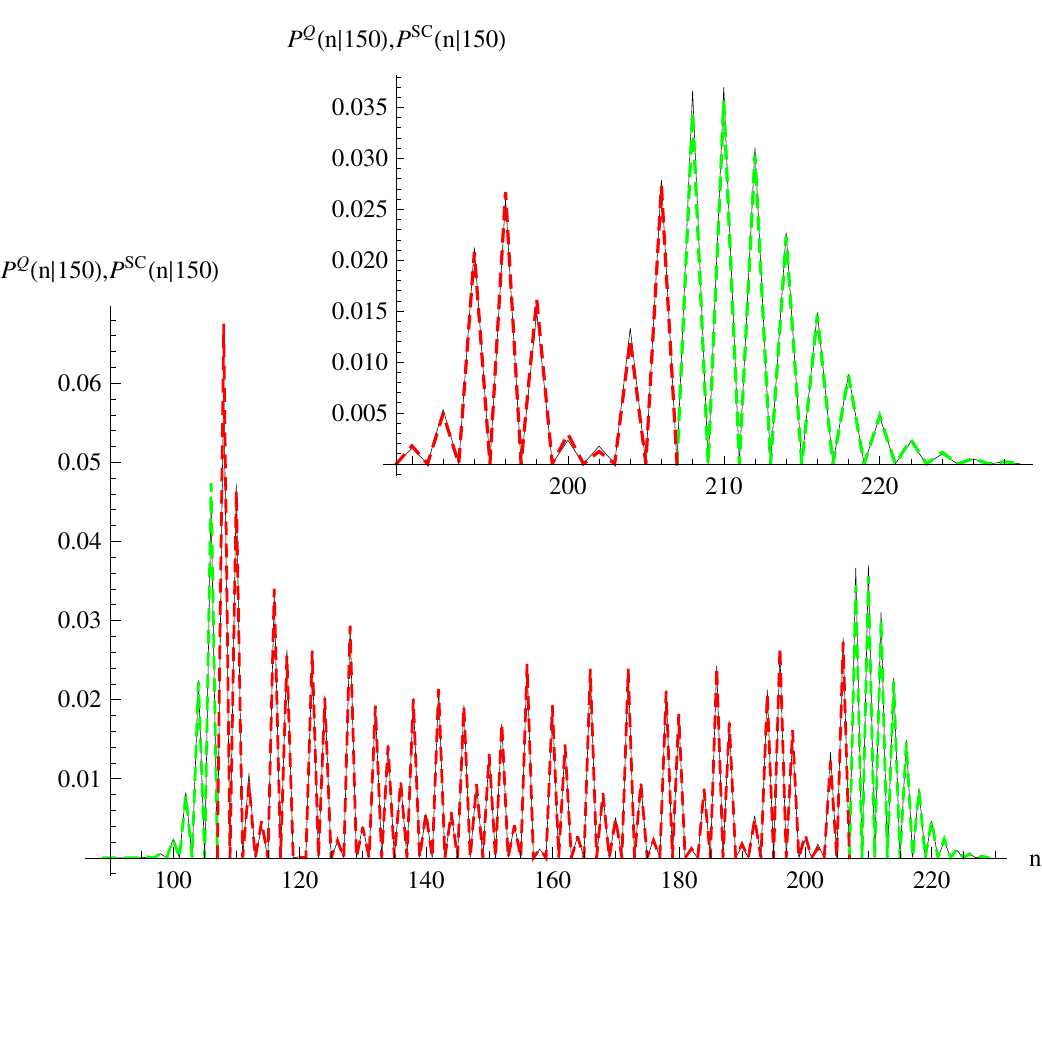}
\caption{Comparison between the quantum transition probabilities evaluated by numerical integration of the Schr\" odinger equation (black, solid line), and the semiclassical approximations given by Eq.~\ref{sc} in the central region $110<n<200$ (red, dashed), and by Eq.~\ref{eq:airy_interference} in the boundary regions $90<n<110$ and $200<n<230$ (green, dashed).}
\label{perfect}
\end{figure}

Let us consider the function
\begin{equation}
\label{eq:DeltaSqe}
\Delta S(q,E) = S^{\cal A}(q) - S^{\cal B}(q,E)
\end{equation}
that appears inside the exponent in Eq.~\ref{eq:innerProduct}, and its derivative
\begin{equation}
\label{eq:Deltapdef}
\Delta p(q,E) = p^{\cal A}(q) - p^{\cal B}(q,E) = \frac{\partial}{\partial q} \Delta S(q,E) \quad .
\end{equation}
We have suppressed the subscripts $b$ and $b^\prime$, and we have explicitly indicated that $S^{\cal B}$ and $p^{\cal B}$ depend on the final energy shell $E$.
In Sec.~\ref{sec:semiclassical} we evaluated the integral $\int {\rm d}q \, \exp(i\Delta S/\hbar)$ by performing a quadratic expansion of $\Delta S$ around a single intersection point $q_l$ between the curves ${\cal A}$ and ${\cal B}$, where $\Delta p$ vanishes (Fig.~\ref{fig:intersection}).
As the energy $E$ approaches $E_{\rm max}$ from below, two points of intersection between ${\cal A}$ and ${\cal B}$ coalesce at a point $(q_c,p_c)$, indicated in Fig.~\ref{fig:doubleWell_noSym}.
Fig.~\ref{fig:coalescence} illustrates the behavior of $\Delta p(q,E)$ and $\Delta S(q,E)$ in the vicinity of $q_c$ and $E_{\rm max}$.
In particular, $\Delta S$ has two stationary points when $E<E_{\rm max}$ and none when $E>E_{\rm max}$.
The stationary phase integration performed in Sec.~\ref{sec:semiclassical} (Eqs.~\ref{eq:SPcondition} - \ref{eq:stationaryPhaseIntegral}) implicitly assumed well-isolated stationary points, but this assumption breaks down near $E=E_{\max}$.
In this situation we evaluate the integral $\int {\rm d}q \, \exp(i\Delta S/\hbar)$ by expanding $\Delta S$ to cubic rather than quadratic order.
Leaving the details of the calculation to Appendix \ref{B}, here we present the result:
\begin{equation}
\label{eq:airyIntegral}
\int {\rm d}q \, \exp\left[ \frac{i}{\hbar} \Delta S(q,E) \right] =
2\pi \left(\frac{2\hbar}{\vert k\vert}\right)^{1/3} \exp\left[\frac{i}{\hbar}\Delta S(q_c,E) \right] {\rm Ai}\left[ - \frac{2^{1/3}\nu}{k^{1/3}\hbar^{2/3}}(E-E_{\rm max})  \right]
\quad ,
\end{equation}
where
\begin{equation}
\label{eq:defknu}
k = \frac{\partial^2\Delta p}{\partial q^2} (q_c,E_{\rm max})
\quad,\quad
\nu = \frac{\partial p^{\cal B}}{\partial E}(q_c,E_{\rm max})
\end{equation}
and ${\rm Ai}(\cdot)$ denotes the Airy function.
Combining this result with Eq.~\ref{eq:innerProduct}, we obtain the following expression, which is valid when $E_n^B \approx E_{\rm max}$:
\begin{equation}
\label{eq:airy_single}
P_1^{SC,tail}(n\vert m) = \left\vert \left\langle \phi_n^B \vert  \psi_\tau \right\rangle \right\vert^2 =
4\pi^2 \, \rho_c^{\cal A} \rho_c^{\cal B}
\left(\frac{2\hbar}{\vert k\vert}\right)^{2/3}
\left\vert {\rm Ai} \left[ - \frac{2^{1/3}\nu}{k^{1/3}\hbar^{2/3}}(E_n^B-E_{\rm max})  \right] \right\vert^2 \quad,
\end{equation}
where $\rho_c^{\cal A,B} = \rho^{\cal A,B}(q_c)$, and the subscript $1$ emphasizes that we have assumed only a single point of tangency between ${\cal A}$ and ${\cal B}$.
At the lower boundary of the classically allowed region, we obtain the same result but with $E_{\rm max}$ replaced by $E_{\rm min}$ in Eqs.~\ref{eq:defknu} and \ref{eq:airy_single}.

The Airy function decays rapidly for positive values of its argument, and oscillates for negative values:
\begin{equation}
\label{eq:airy_asymp}
{\rm Ai}(\zeta) \sim \frac{1}{2\sqrt{\pi}} \zeta^{-1/4} \exp\left( -\frac{2}{3}\zeta^{3/2} \right)
\quad,\quad
{\rm Ai}(-\zeta) \sim \frac{1}{\sqrt{\pi}} \zeta^{-1/4} \sin\left( \frac{2}{3}\zeta^{3/2} + \frac{\pi}{4} \right)
\end{equation}
for real $\zeta\gg 1$.
One can establish by inspection that $\nu>0>k$ when $E>E_{\rm max}$ and $\nu,k>0$ when $E<E_{\rm min}$, therefore the argument of the Airy function in $P_1^{SC,tail}$ is positive outside the classically allowed region.
As a result the transition probability decays monotonically, penetrating the classically forbidden region over a characteristic skin depth $\vert k^{1/3}\hbar^{2/3}\nu^{-1}\vert$.

Let us now consider the situation corresponding to the example of Sec.~\ref{sec:numerical}, in which there are simultaneously two points of tangency between the curves ${\cal A}$ and ${\cal B}$ at $E=E_{\rm max}$, and these two points are related by symmetry (Fig.~\ref{fig:quartic_Sym}).
Using the subscripts $c$ and $d$ to distinguish these two points, we have:
\begin{equation}
(q_c,p_c) = (-q_d,-p_d)
\quad,\quad
\nu_c = -\nu_d \equiv \nu
\quad,\quad
k_c = -k_d \equiv k
\quad,\quad
\rho_c^{\cal A,\cal B} = \rho_d^{\cal A,\cal B} \quad.
\end{equation}
Summing the contributions from these two points gives us
\begin{equation}
\label{eq:innerProduct_2coalesc}
\left\langle \phi_n^B \vert  \psi_\tau \right\rangle =
2\pi \sqrt{\rho_c^{\cal A} \rho_c^{\cal B}} \left(\frac{2\hbar}{\vert k\vert}\right)^{1/3}
{\rm Ai}\left[ - \frac{2^{1/3}\nu}{k^{1/3}\hbar^{2/3}}(E-E_{\rm max})  \right]
\left[
e^{i(\Delta S_c/\hbar - \Delta\mu_c \pi/2)} +
e^{i(\Delta S_d/\hbar - \Delta\mu_d \pi/2)}
\right]
\end{equation}
where $\Delta S_{c,d}$ are evaluated at $E=E_{\rm max}$.
Following arguments similar to those leading to Eq.~\ref{eq:fastInterference}, we obtain
\begin{equation}
\label{eq:phases}
e^{i(\Delta S_c/\hbar - \Delta\mu_c \pi/2)} + e^{i(\Delta S_d/\hbar - \Delta\mu_d \pi/2)} =
e^{i(\Delta S_c/\hbar - \Delta\mu_c \pi/2)} \left[ 1 + e^{i(m-n)\pi} \right] \quad.
\end{equation}
Eqs.~\ref{eq:innerProduct_2coalesc} and \ref{eq:phases} lead to a result identical to Eq.~\ref{eq:airy_single}, apart from a factor that captures the interference between the two parity-related trajectories:
\begin{equation}
\label{eq:airy_interference}
P_2^{SC,tail}(n\vert m) = P_1^{SC,tail}(n\vert m)  \cdot
4 \cos^2\left[ \frac{(m-n)\pi}2 \right] \quad ,
\end{equation}
with the subscript indicating two points of tangency between ${\cal A}$ and ${\cal B}$.

Fig.~\ref{perfect} compares the quantum transition probability $P^Q$ with our semiclassical approximations given by $P^{SC}$ (Eq.~\ref{sc}) and $P_2^{SC,tail}$ (Eq.~\ref{eq:airy_interference}), in the central and boundary regions, respectively.
The agreement is excellent, illustrating that the entire quantum work distribution can be understood in terms of contributions from individual classical trajectories, and the interferences between them.

\section{discussion and conclusion}

The {\it correspondence principle} broadly states that classical mechanics is recovered from quantum mechanics in the limit of large quantum numbers.
In textbook discussions this principle is often illustrated in terms of static properties, typically by comparing the coordinate space probability distribution of a harmonic oscillator eigenstate with the classical, microcanonical  distribution at the same energy~\cite{Schiff1968,Griffiths1995}.
Here, by contrast, we have considered the correspondence principle in a dynamic setting, comparing classical and quantum transition probabilities under a time-dependent Hamiltonian. In this setting, time-dependent WKB theory provides a set of tools for using classical trajectories to construct approximate solutions of quantum dynamics.

We have applied these tools to study the relationship between the definitions of quantum and classical {\it work} given by Eqs.~\ref{eq:qwork} and \ref{eq:cwork}.
Focusing on the transition probabilities $P^Q(n\vert m)$ and $P^C(n\vert m)$ for systems with one degree of freedom, we have obtained three main conclusions, and have illustrated them with simulations of a driven quartic oscillator:
\begin{itemize}
\item
In the classically allowed region, the quantum transition probability can be approximated in terms of interfering classical trajectories (Eq.~\ref{sc}, Fig.~\ref{sctransitionprobabilities}).
\item
When the interferences between these trajectories are ignored, the classical transition probability is obtained (Eq.~\ref{eq:Pnm_result}, Fig.~\ref{transitionprobabilities}).
\item
The tunneling of the quantum transition probability into the classically forbidden region is accurately described by a decaying Airy function, whose arguments are expressed in terms of classical quantities (Eqs.~\ref{eq:airy_single}, \ref{eq:airy_interference}, Fig.~\ref{perfect}).
\end{itemize}
The second conclusion, and Fig.~\ref{transitionprobabilities} in particular, clarifies the sense in which classical mechanics is ``recovered'' in the limit of large quantum numbers (in this context), whereas the first and third conclusions describe inherently quantal effects -- interference, tunneling -- that can nevertheless be understood and approximated in terms of classical trajectories.
In view of similar analyses in atomic and molecular contexts (see e.g.\ Refs.~\cite{Marcus1970,Marcus1971,Miller1974,Schwieters1995a,Schwieters1995b}) these conclusions are not surprising.
However, given recent interest in quantum nonequilibrium work relations, our results provide a timely investigation of the correspondence principle as it applies to the {\it work} performed on a system driven away from equilibrium.
As we have shown, semiclassical mechanics provides a bridge between classical and quantal work distributions, and our conclusions provide some justification for the two-point measurement definition of quantum work (Eq.~\ref{eq:qwork}).

It will be interesting to study whether our results generalize to systems with more than one degree of freedom.
For $N$-particle systems with $N\gg 1$, we generically expect that the classical transition probability $P^C(n\vert m)$ will be bell-shaped rather than U-shaped, with $n_{\rm min}$ and $n_{\rm max}$ found deep in the tails of $P^C(n\vert m)$.
In this situation the quantum tunneling into the forbidden regions will be negligible, since both $P^C(n\vert m)$ and $P^Q(n\vert m)$ will be very small at $n_{\rm min}$ and $n_{\rm max}$.
An obstacle to studying $N$-particle systems analytically
is the lack of semiclassical representations of energy eigenstates, analogous to Eq.~\ref{eq:eigenstate}, for generic systems with multiple degrees of freedom.
Progress might be made by studying two limiting cases: {\it integrable} and fully {\it chaotic} systems \cite{Berry1983}.
For systems whose classical dynamics are integrable, semiclassical expressions for energy eigenstates can be constructed using classical action-angle variables (see e.g. Eq.~11 of Ref.~\cite{VanVleck1928} or Eq.~3.5 of Ref.~\cite{Berry1977a} or Eq.~37 of Ref.~\cite{Berry1979}).
At the other extreme, for fully chaotic systems, {\it Berry's conjecture}~\cite{Berry1977b} suggests that energy eigenstates can be treated as Gaussian random functions of the configuration, ${\bf q}$.
In either case, the expression for the energy eigenstate might be combined with time-dependent WKB theory as a first step toward generalizing Eq.~\ref{eq:innerProduct} and subsequent results to multi-dimensional systems.

\acknowledgments
HTQ acknowledges Meng-Li Du for helpful discussions. CJ gratefully acknowledges support from the National Science Foundation (USA) under grants DMR-1206971 and DMR-1506969. HTQ gratefully acknowledges support from the National Science Foundation of China under grant 11375012, and The Recruitment Program of Global Youth Experts of China. SR and CJ are grateful for support from the US-Israel Binational Science Foundation (grant 2010363). SR gratefully acknowledges the support of the Israel Science Foundation (grant 924/11). This work was partially supported by the COST Action MP1209.

\appendix

\section{\label{A}Brief review of time-dependent WKB theory}

In time-dependent WKB theory, approximate solutions of the time-dependent Schr\" odinger equation are expressed in terms of classical constructions in phase space.
In the brief summary that follows, we restrict ourselves to a single degree of freedom, and a Hamiltonian of the form $H(q,p,t) = p^2/2M + V(q,t)$. 

A wavefunction $\psi(q,t)$ is said to be in WKB form when written as the product of a slowly varying amplitude and a rapidly oscillating phase,
\begin{equation}
\label{eq:WKBform}
\psi(q,t) = A(q,t) \exp\left[ \frac{i}{\hbar} S(q,t) \right]
\quad ,
\end{equation}
or else as a sum of such terms, as in Eq.~(\ref{eq:WKBsoln}) below.
Locally, Eq.~(\ref{eq:WKBform}) describes a wave train $\psi \propto \exp(ikq)$, with wave number $k = \hbar^{-1} \partial S/\partial q$.
This implies a local momentum
\begin{equation}
\label{eq:momentumField}
p(q,t) = \hbar k = \frac{\partial S}{\partial q}(q,t)
\quad,
\end{equation}
and the Born rule implies a probability density
\begin{equation}
\label{eq:density}
\rho(q,t) = \vert A(q,t)\vert^2
\quad .
\end{equation}
These functions inherit their time-dependence from $\psi(q,t)$.
A standard calculation, in which Eq.~(\ref{eq:WKBform}) is substituted into the Schr\" odinger equation and $\hbar$ is treated as a small parameter~\cite{Littlejohn1992}, produces the equations of motion
\begin{subequations}
\label{eq:HJ+cont}
\begin{eqnarray}
\label{eq:HJ}
\frac{\partial S}{\partial t} + H\left( q , \frac{\partial S}{\partial q} , t\right) &=&0 \\
\label{eq:cont}
\frac{\partial \rho}{\partial t} + \frac{\partial}{\partial q} \left( \frac{p}{M} \rho \right) &=& 0
\quad ,
\end{eqnarray}
\end{subequations}
where terms of order $\hbar^2$ have been neglected.
Eq.~\ref{eq:HJ} is the Hamilton-Jacobi equation.
Eq.~\ref{eq:cont} is the continuity equation for evolution under a velocity field $p(q,t)/M$.

To interpret these equations, let the function $p(q,0)$ describe a curve ${\cal A}_{0}$ in phase space, at an initial time $t=0$.
If the wavefunction $\psi$ is represented as a sum of terms of the form given by Eq.~\ref{eq:WKBform}, then $p(q,0)$ is multivalued, and the curve ${\cal A}_{0}$ has multiple branches, but for the moment we focus on a single branch.
${\cal A}_{0}$ is a {\it Lagrangian manifold}, determined by its {\it generating function}, $S(q,0)$, via Eq.~\ref{eq:momentumField}.
Now imagine that this curve carries a probability density whose projection onto the coordinate axis is $\rho(q,0)$.
It is convenient to picture a swarm of particles, sprinkled along ${\cal A}_{0}$ so that the fraction of particles between $q$ and $q+{\rm d}q$ is given by  $\rho(q,0) \, {\rm d}q$.
From these initial conditions, each particle in the swarm evolves under Hamilton's equations, giving rise to a time-dependent curve ${\cal A}_{t}$ and density $\rho(q,t)$.
The Hamilton-Jacobi equation governs the evolution of ${\cal A}_{t}$, through its generating function $S(q,t)$, and the continuity equation governs the evolution of the projected density $\rho(q,t)$.

With these considerations in mind, approximate solutions of the time-dependent Schr\" odinger equation are written in the form
\begin{equation}
\label{eq:WKBsoln}
\psi(q,t) = \sum_b \sqrt{\rho_b(q,t)} \exp \left[ \frac{i}{\hbar} S_b(q,t) - i\mu_b \frac{\pi}{2} \right] \quad.
\end{equation}
The sum is taken over the branches of a Lagrangian manifold ${\cal A}_{t}$ whose generating function $S(q,t)$ satisfies the Hamilton-Jacobi equation (Eq.~\ref{eq:HJ}), and the projected probability density for each branch satisfies the continuity equation (Eq.~\ref{eq:cont}).

The quantities $\mu_b$ are {\it Maslov indices}.
These determine the relative quantum phases of the various branches of ${\cal A}_{t}$, and they differ from one another by integer values.
(By convention, the overall phase of $\psi$ is adjusted so that the $\mu_b$'s themselves are integers.)
See Ref.~\cite{Littlejohn1992} for a more detailed discussion of Maslov indices, as well as a careful treatment of {\it caustic points}, where two branches of ${\cal A}_t$ meet and the manifold becomes ``vertical'' in the $(q,p)$-plane.

At the semiclassical level of approximation, Eq.~(\ref{eq:WKBsoln}) connects the evolution of a quantum wavefunction to that of a swarm of classical particles ``surfing'' on a Lagrangian manifold.
This perspective provides a natural starting point for a comparison between quantum and classical work distributions.

As an important example of a WKB wavefunction, consider a manifold ${\cal A}_0$ that is a single energy shell $E$ of a {\it fixed} Hamiltonian $H(q,p)$, and consider a microcanonical distribution of initial conditions on this manifold.
Under time evolution, neither the manifold nor the probability distribution changes, as each trajectory merely goes round the energy shell.
The solution of Eq.~\ref{eq:HJ} in this case is
\begin{equation}
S_b(q,t) = S_b(q,0) - Et \quad,
\end{equation}
hence by Eq.~\ref{eq:WKBsoln} the wavefunction simply acquires a phase $e^{-iEt/\hbar}$.
Thus in the semiclassical limit, energy eigenstates correspond to microcanonical ensembles.
Indeed, Eq.~\ref{eq:WKBsoln} in this case leads to the WKB approximation for energy eigenstates that is familiar from undergraduate textbooks on quantum mechanics, where it is typically derived in a different manner.
Moreover, a proper treatment of the Maslov indices leads to the quantization condition $\oint p(q)\, {\rm d}q = [m + (1/2)] h$.

\section{\label{B}Semiclassical transition probabilities in the boundary area}

Expanding the function $\Delta S(q,E)$ defined by Eq.~\ref{eq:DeltaSqe} to cubic order, we get
\begin{equation}
\label{eq:cubicExpansion}
\Delta S(q,E) \approx \Delta S_c(E) + \Delta S_c^\prime(E) \, (q-q_c) + \frac{1}{2} \Delta S_c^{\prime\prime}(E) \, (q-q_c)^2 + \frac{1}{6} \Delta S_c^{\prime\prime\prime}(E) \, (q-q_c)^3 \quad ,
\end{equation}
where $\Delta S_c(E) = \Delta S(q_c,E)$ and primes indicate derivatives with respect to $q$. Note that here we expand $\Delta S$ around the coalescence point $q=q_c$, where the surface ${\cal A}$ becomes tangent to the surface ${\cal B}$ (see Fig.~\ref{fig:coalescence}).
We now wish to evaluate the integral $\int {\rm d}q \, \exp(i\Delta S/\hbar)$, using Eq.~\ref{eq:cubicExpansion}.

Let us first consider the integral
\begin{equation}
\label{eq:basicIntegral}
I = \int_{-\infty}^{+\infty} {\rm d}q \, \exp\left[ \frac{i}{\hbar} f(q) \right]
\quad,\quad
f(q) = \alpha_0 + \alpha_1 q + \frac{1}{2} \alpha_2 q^2 + \frac{1}{6} \alpha_3 q^3 \quad .
\end{equation}
Rewriting $f(q)$ as
\begin{equation}
f(q) = \beta_0 + \beta_1 (q-c) + \frac{1}{3} \beta_3 (q-c)^3 \quad,
\end{equation}
where
\begin{equation}
\label{eq:alphabeta}
c = -\frac{\alpha_2}{\alpha_3} \quad , \quad
\beta_0 = \alpha_0 - \frac{\alpha_1\alpha_2}{\alpha_3} +  \frac{\alpha_2^3}{3\alpha_3^2} \quad , \quad
\beta_1 = \alpha_1 - \frac{\alpha_2^2}{2\alpha_3} \quad , \quad
\beta_3 = \frac{\alpha_3}{2} \quad ,
\end{equation}
we obtain
\begin{equation}
\label{eq:basicIntegral_result}
I = e^{i\beta_0/\hbar} \int_{-\infty}^{+\infty} {\rm d}y \, \exp\left[ \frac{i}{\hbar} \left(\beta_1 y + \frac{\beta_3}{3} y^3\right) \right]
=2\pi e^{i\beta_0/\hbar}\left(\frac{\hbar}{\vert\beta_3\vert}\right)^{1/3}  {\rm Ai}\left( \frac{\beta_1}{\hbar^{2/3}\beta_3^{1/3}} \right)
\end{equation}
using the integral representation of the Airy function,
\begin{equation}
{\rm Ai}(\zeta) = \frac{1}{2\pi}  \int_{-\infty}^{+\infty} {\rm d}t \, \exp\left[ i\left( \zeta t + \frac{t^3}{3} \right) \right]
\end{equation}
We will now use this result to evaluate
\begin{equation}
\label{eq:integral}
I(E) = \int {\rm d}q \, \exp\left[ \frac{i}{\hbar} \Delta S(q,E) \right]
\end{equation}
for small $\hbar$, applying the stationary phase approximation and treating $E$ as a parameter of the integral.

The cubic expansion for $\Delta S$ given by Eq.~\ref{eq:cubicExpansion} gives us the coefficients
\begin{eqnarray}
\alpha_0(E) &=& \Delta S(q_c,E) \\
\alpha_1(E) &=& \Delta S^\prime(q_c,E) = \Delta p(q_c,E) \\
\alpha_2(E) &=& \Delta S^{\prime\prime}(q_c,E) = \Delta p^\prime(q_c,E) \\
\alpha_3(E) &=& \Delta S^{\prime\prime\prime}(q_c,E) = \Delta p^{\prime\prime}(q_c,E)
\end{eqnarray}
with $\Delta p = p^{\cal A} - p^{\cal B} = \Delta S^\prime$ (Eq.~\ref{eq:Deltapdef}).
Note that
\begin{equation}
\alpha_1(E_c) = \alpha_2(E_c)  = 0\quad ,
\end{equation}
since ${\cal A}$ and ${\cal B}$ are tangent at $q_c$ when $E=E_c$.
The integral $I(E)$ is now given by Eq.~(\ref{eq:basicIntegral_result}), with $\beta_0$, $\beta_1$ and $\beta_3$ obtained from $\alpha_0$, $\alpha_1$, $\alpha_2$ and $\alpha_3$ via Eq.~(\ref{eq:alphabeta}).
This result simplifies if we consider how the $\beta$'s behave in the vicinity of $E=E_c$ as $\hbar\rightarrow 0$.

Let us define
\begin{eqnarray}
\label{eq:usefulConstants}
p_c &=& p^{\cal A}(q_c) = p^{\cal B}(q_c,E_c) \\
p_c^\prime &=& \frac{\partial p^{\cal A}}{\partial q} (q_c) = \frac{\partial p^{\cal B}}{\partial q}(q_c,E_c) \\
k &=& \frac{\partial^2 p^{\cal A}}{\partial q^2}(q_c) - \frac{\partial^2 p^{\cal B}}{\partial q^2}(q_c,E_c) = \alpha_3(E_c)  \\
\nu &=& \frac{\partial p^{\cal B}}{\partial E}(q_c,E_c)
\end{eqnarray}
Now look at the argument of the Airy function in Eq.~(\ref{eq:basicIntegral_result}),
\begin{equation}
\zeta(E,\hbar) = \frac{\beta_1(E)}{\hbar^{2/3}\beta_3(E)^{1/3}}
\end{equation}
and consider a fixed range of $\zeta$ values, $\zeta_{-}<0<\zeta_{+}$, chosen so that the asymptotic approximations for the Airy function (Eq.~\ref{eq:airy_asymp}) are accurate at both $\zeta_{-}$ and $\zeta_{+}$.
Noting that $\zeta(E_c,\hbar) = 0$ and expanding $\zeta(E,\hbar)$ to first order in $(E-E_c)$, we see that as $\hbar\rightarrow 0$, the fixed range $[\zeta_{-},\zeta_{+}]$ translates to a range of energies whose width scales like $\hbar^{2/3}$.
Thus we will be interested in energies that have the following scaling relation:
\begin{equation}
\label{eq:scaling}
\epsilon \sim \hbar^{2/3} \quad ,
\end{equation}
where $\epsilon = E-E_c$.
To leading order in $\hbar$, and expressing quantities in terms of $\epsilon$ rather than $E$, we  have
\begin{equation}
\zeta = \frac{a\epsilon}{\hbar^{2/3}b^{1/3}}
\quad,\quad {\rm where} \quad
a = \frac{\partial\beta_1}{\partial\epsilon}(\epsilon=0) \quad,\quad b = \beta_3(\epsilon=0) \quad.
\end{equation}
Now recall that $\beta_1 = \alpha_1 - \alpha_2^2/2\alpha_3$.
In the vicinity of $\epsilon=0$ (where $\alpha_1=\alpha_2=0$), we have $\alpha_1\sim\epsilon\sim\hbar^{2/3}$ and $\alpha_2^2/2\alpha_3\sim\epsilon^2\sim\hbar^{4/3}$, hence the latter term can be ignored.
We thus write
\begin{equation}
a = \frac{\partial\alpha_1}{\partial\epsilon}(0) = -\nu
\quad,\quad
b = \frac{1}{2} \alpha_3(0) = \frac{k}{2} \quad,
\end{equation}
therefore
\begin{equation}
\zeta = - \left(\frac{2}{k\hbar^2}\right)^{1/3} \nu\epsilon \quad.
\end{equation}
Combining results and discarding terms that vanish as $\hbar\rightarrow 0$, Eq.~(\ref{eq:basicIntegral_result}) finally gives us
\begin{equation}
I(\epsilon) = 2\pi \left(\frac{2\hbar}{\vert k\vert}\right)^{1/3} \exp\left[\frac{i}{\hbar}\alpha_0(\epsilon)\right] {\rm Ai}\left( - \frac{2^{1/3}\nu\epsilon}{k^{1/3}\hbar^{2/3}}  \right)
\quad ,
\label{phaseintegral}
\end{equation}
which is equivalent to Eq.~\ref{eq:airyIntegral}.

\bibliography{correspondence_principle}

\end{document}